\documentclass[acmsmall, screen, authorversion, nonacm,review=false,timestamp=false]{acmart}

\usepackage[horizontal, grid=lightgray, skipempty]{credits}
\usepackage[capitalize]{cleveref}
\usepackage[inline]{enumitem}
\usepackage{comment}
\usepackage[nolist]{acronym}
\usepackage{caption}
\usepackage{subcaption}
\usepackage{tabularx}
\usepackage{array}
\usepackage{booktabs}
\usepackage{longtable}
\usepackage{xcolor}
\usepackage[font=small,skip=1pt]{caption}

\newcommand{\revi}[1]{#1}

\setcopyright{acmcopyright}
\copyrightyear{2025}
\acmYear{2025}
\acmDOI{XXXXXXX.XXXXXXX}

\acmJournal{CSUR}
\acmVolume{37}
\acmNumber{4}
\acmMonth{8}

\acmPrice{15.00}
\acmISBN{978-1-4503-XXXX-X/18/06}

\begin{document}

\title{Explanation User Interfaces: A Systematic Literature Review}
\titlenote{This research is partially funded by \begin{enumerate*}[label={(\roman*)}]
    \item the \grantsponsor{DevProDev}{Italian Ministry of University and Research (MUR) and by the European Union - NextGenerationEU}, Mission 4, Component 2, Investment 1.1, under grant PRIN 2022 ``DevProDev: Profiling Software Developers for Developer-Centered Recommender Systems''---\grantnum{DevProDev}{CUP: H53D23003620006};
    \item the \grantsponsor{PROTECT}{Italian Ministry of University and Research (MUR) and by the European Union - NextGenerationEU}, Mission 4, Component 2, Investment 1.1, under grant PRIN 2022 PNRR ``PROTECT: imPROving ciTizEn inClusiveness Through Conversational AI'' \grantnum{PROTECT}{(Grant P2022JJPBY)---CUP: H53D23008150001}.
    \item the co-funding of the \grantsponsor{FAIR6}{European Union---Next Generation EU}: NRRP Initiative, Mission 4, Component 2, Investment 1.3 -- Partnerships extended to universities, research centres, companies, and research D.D. MUR n. 341 del 15.03.2022 -- Next Generation EU (PE0000013 -- ``Future Artificial Intelligence Research - FAIR'' - Spoke 6 ``Symbiotic AI''---\grantnum{FAIR6}{CUP: H97G22000210007}).
    \item the Partnership Extended \grantnum{FAIR1}{PE00000013} - \grantsponsor{FAIR1}{``FAIR - Future Artificial Intelligence Research'' - Spoke 1 ``Human-centered AI''} 
    \item \grantsponsor{TANGO}{TANGO project}, grant agreement no. \grantnum{TANGO}{101120763},
    \item The \grantsponsor{HORIZON}{European Community Horizon~2020} programme under the funding scheme \grantnum{HORIZON}{ERC-2018-ADG G.A. 834756} \textit{XAI: Science and technology for the eXplanation of AI decision making}
\end{enumerate*}The research of Andrea Esposito is funded by a Ph.D. fellowship within the framework of the \grantsponsor{DM352AE}{Italian ``D.M. n. 352, April 9, 2022''} - under the National Recovery and Resilience Plan, Mission 4, Component 2, Investment 3.3 - Ph.D. Project ``Human-Centered Artificial Intelligence (HCAI) techniques for supporting end users interacting with AI systems,'' co-supported by ``Eusoft S.r.l.'' (\grantnum{DM352AE}{CUP H91I22000410007}). The research of Francesco Greco is funded by a Ph.D. fellowship within the framework of the \grantsponsor{DM352FG}{Italian ``D.M. n. 352, April 9, 2022''} - under the National Recovery and Resilience Plan, Mission 4, Component 2, Investment 3.3 - Ph.D. Project ``Investigating XAI techniques to help users defend from phishing attacks'', co-supported by ``Auriga S.p.A.'' (\grantnum{DM352FG}{CUP H91I22000410007}).}
\author{Eleonora Cappuccio}
\affiliation{%
  \institution{University of Pisa}
  \department{Department of Computer Science}
  \streetaddress{Largo B. Ponte Corvo 3}
  \city{Pisa}
  \country{Italy}
  \postcode{56127}
}
\affiliation{%
  \institution{University of Bari Aldo Moro}
  \department{Department of Computer Science}
  \streetaddress{Via E. Orabona 4}
  \city{Bari}
  \country{Italy}
  \postcode{70125}
}
\affiliation{%
  \institution{ISTI CNR}
  \streetaddress{Via G. Moruzzi 1}
  \city{Pisa}
  \country{Italy}
  \postcode{56124}
}
\email{eleonora.cappuccio@cnr.it}
\orcid{0000-0002-6105-2512}
\authornote{Corresponding author.}

\author{Andrea Esposito}
\affiliation{%
  \institution{University of Bari Aldo Moro}
  \department{Department of Computer Science}
  \streetaddress{Via E. Orabona 4}
  \city{Bari}
  \country{Italy}
  \postcode{70125}
}
\email{andrea.esposito@uniba.it}
\orcid{0000-0002-9536-3087}

\author{Francesco Greco}
\affiliation{%
  \institution{University of Bari Aldo Moro}
  \department{Department of Computer Science}
  \streetaddress{Via E. Orabona 4}
  \city{Bari}
  \country{Italy}
  \postcode{70125}
}
\email{francesco.greco@uniba.it}
\orcid{0000-0003-2730-7697}

\author{Giuseppe Desolda}
\affiliation{%
  \institution{University of Bari Aldo Moro}
  \department{Department of Computer Science}
  \streetaddress{Via E. Orabona 4}
  \city{Bari}
  \country{Italy}
  \postcode{70125}
}
\email{giuseppe.desolda@uniba.it}
\orcid{0000-0001-9894-2116}

\author{Rosa Lanzilotti}
\affiliation{%
  \institution{University of Bari Aldo Moro}
  \department{Department of Computer Science}
  \streetaddress{Via E. Orabona 4}
  \city{Bari}
  \country{Italy}
  \postcode{70125}
}
\email{rosa.lanzilotti@uniba.it}
\orcid{0000-0002-2039-8162}

\author{Salvatore Rinzivillo}
\affiliation{%
  \institution{ISTI CNR}
  \streetaddress{Via G. Moruzzi 1}
  \city{Pisa}
  \country{Italy}
  \postcode{56124}
}
\email{rinzivillo@isti.cnr.it}
\orcid{0000-0003-4404-4147}

\renewcommand{\shortauthors}{Cappuccio et al.}

\begin{abstract}
  Artificial Intelligence (AI) is one of the major technological advancements of this century, bearing incredible potential for users through AI-powered applications and tools in numerous domains. Being often black-box (i.e., its decision-making process is unintelligible), developers typically resort to eXplainable Artificial Intelligence (XAI) techniques to interpret the behaviour of AI models to produce systems that are transparent, fair, reliable, and trustworthy. However, presenting explanations to the user is not trivial and is often left as a secondary aspect of the system's design process, leading to AI systems that are not useful to end-users. This paper presents a Systematic Literature Review on Explanation User Interfaces (XUIs) to gain a deeper understanding of the solutions and design guidelines employed in the academic literature to effectively present explanations to users. To improve the contribution and real-world impact of this survey, we also present a \revi{platform to support} Human-cEnteRed developMent of Explainable user interfaceS (HERMES) and guide practitioners and \revi{scholars} in the design and evaluation of XUIs.
\end{abstract}

\ccsdesc[500]{General and reference~Surveys and overviews}
\ccsdesc[500]{Human-centered computing~User interface design}
\ccsdesc[300]{Computing methodologies~Artificial intelligence}

\keywords{Explainable Artificial Intelligence (AI), XUI, design, evaluation, visualization, explanation}

\received{21 May 2025}
\received[revised]{22 Jan 2026}

\begin{acronym}[HCXAI]
    \acro{HCXAI}{Human-Centered eXplainable Artificial Intelligence}
    \acro{GDPR}{General Data Protection Regulations}
    \acro{LLM}{Large Language Model}
    \acro{UCD}{User-Centered Design}
    \acro{ML}{Machine Learning}
    \acro{DL}{Deep Learning}
    \acro{XUI}{Explanation User Interface}
    \acro{HCAI}{Human-Centered Artificial Intelligence}
    \acro{XAI}{eXplainable Artificial Intelligence}
    \acro{AI}{Artificial Intelligence}
    \acro{HCD}{Human-Centered Design}
    \acro{HCI}{Human-Computer Interaction}
    \acro{UI}{User Interface}
    \acro{SLR}{Systematic Literature Review}
    \acro{EU}{European Union}
    \acro{LIME}{Locally Interpretable Model-agnostic Explanations}
    \acro{SHAP}{SHapley Additive exPlanation}
    \acro{CAM}{Class Activation Mapping}
    \acro{Grad-CAM}{Gradient-weighted Class Activation Mapping}
    \acro{PDP}{Partial Dependence Plot}
    \acro{DiCE}{Diverse Counterfactual Explanations}
    \acro{DARPA}{Defense Advanced Research Projects Agency}
    \acro{HERMES}{a framework for Human-cEnteRed developMent of Explainable user interfaceS}
\end{acronym}
\maketitle

\section{Introduction}

Recent years have witnessed a remarkable interest in the field of \ac{AI}, which has become a prominent topic in both academia and industry. The rapid advancements in \ac{AI} technologies have led to their widespread adoption across various sectors, including healthcare, finance, and transportation. The \acl{ML} subfield of \ac{AI} has gained particular attention due to its ability to provide discriminative and predictive capabilities, enabling the development of sophisticated applications such as image recognition, natural language processing, and autonomous systems. The increasing complexity of \acl{ML} models has raised concerns about their interpretability and transparency, particularly for those models based on deep learning architectures. These models often operate as ``black boxes,'' making it challenging for users to understand their decision-making processes~\cite{Guidotti2019Survey}. This has led to a growing demand for \ac{XAI} techniques that can elucidate the decision-making processes of these models. Black-box \acl{ML} models pose various threats to their users. For example, the risk of unfair and unjust bias increases when the model's inner workings are not clear \cite{Mehrabi2022Survey}. Therefore, worldwide legislative bodies are attempting to regulate the use of such systems. For example, the \ac{EU} recently released the \ac{AI} act with this goal \cite{EuropeanParliament2024Regulation}. Most of these regulations advocate the need for \ac{AI} systems to be accountable, fair, and \emph{explainable} \cite{EuropeanParliament2024Regulation}. In this context, \ac{XAI} may be beneficial. \ac{XAI} is a sub-field of \ac{AI}, initially formalised by the \ac{DARPA} in 2017 \cite{Gunning2019DARPAs}, which focuses on ``opening'' black-box models, making their inner workings clear, explainable, and interpretable \cite{Guidotti2019Survey,Gunning2019DARPAs}. Although different techniques for explaining black-box models exist (as will be detailed in \Cref{sec:xai}), a common and precise definition of ``good'' explanations has yet to be established. Several studies have pointed out that the majority of the work on \ac{XAI} is based on researchers' intuition of what qualifies as a ``good'' explanation \cite{Miller2019Explanation, DBLP:conf/chi/AbdulVWLK18, DBLP:journals/corr/abs-2110-10790liao, DBLP:journals/cacm/Lipton18, DBLP:conf/chi/ChengWZOGHZ19}, effectively framing \ac{XAI} primarily as an algorithmic problem. As a result, there is a gap between \ac{XAI} algorithms designed by researchers and their deployment in real-word scenarios~ \cite{DBLP:journals/corr/abs-2110-10790liao}. To fill this gap, \ac{HCI} research has recently begun to focus on AI-enabled applications and \ac{XAI}, as testified by the birth of the research field of \ac{HCAI}, which reframes AI-development in a human-centred perspective by focusing on users' needs \cite{Shneiderman2022HumanCentered}. Since the first conceptualisation of \ac{XAI} by \ac{DARPA}, the design of \ac{XAI} systems followed two different stages: the design of the algorithmic model itself, and the design of the \ac{UI} \cite{Chromik2021HumanXAI}. Such \acp{UI} are usually referred to as \acp{XUI} \cite{Chromik2021HumanXAI}. As with other types of \acp{UI}, \acp{XUI} must be designed following a human-centred approach to be successful and usable \cite{ISO2018924111}.

In this study, we present a \ac{SLR}, following \citeauthor{Kitchenham2004Procedures}'s procedure \cite{Kitchenham2004Procedures}, that aims at providing a comprehensive overview of the design of \acp{XUI}. Our overall goal is to identify design trends among human-centred \acp{XUI} and \ac{XAI} algorithms to provide a cohesive understanding of the \ac{HCXAI} landscape. More precisely, we contribute to the state-of-the-art systematising the knowledge on four main interrelated aspects of \acp{XUI} design;
\begin{enumerate*}[label=(\roman*)]
    \item The \ac{XAI} models used to generate the explanations,
    \item The solutions adopted to provide the explanations,
    \item The user studies and human-centred techniques used to evaluate the designs,
    \item The design guidelines and frameworks that can help designers and practitioners.
\end{enumerate*} This article is structured as follows: \Cref{sec:background} presents key background concepts; \Cref{sec:methods} presents the methodology used to conduct the \ac{SLR} and the four research questions that guided it; 
\revi{\Cref{sec:RQ1,sec:RQ2,sec:RQ3,sec:RQ4}}report the results to answer each of the four research questions;
\revi{\Cref{sec:hermes}} presents a set of conclusions drawn by analyzing the dimensions of each RQ; moreover, it presents HERMES, \revi{a platform} to guide practitioners in the design and evaluation of XUIs; \revi{\Cref{sec:challenges}} reports open research challenges; \revi{\Cref{sec:limitations}} presents the limitations of our study, highlighting any potential threat to its validity; finally, \revi{\Cref{sec:conclusions}} concludes the article.

\section{Background} \label{sec:background}

This section \revi{provides an overview of the background of our work}. It provides an overview of the field of \ac{XAI}, reprising a classification of methods already available in the literature that can guide in understanding the results of this \ac{SLR}. Furthermore, it explores the need for a human-centred process and its implications for \acp{XUI}.

\subsection{Explainable Artificial Intelligence}\label{sec:xai}

\revi{\Acf{AI} is a field of computer science that is focused on methods for building systems that perform tasks associated with human intelligence, such as reasoning and problem-solving~\cite{Russell2016Artificial}. \acf{ML}, a subset of \acs{AI}, enables systems to learn patterns and make predictions from data without explicit programming, using techniques such as neural networks~\cite{Mitchell1997Machine}. \Acf{DL}, a further specialization of \ac{ML}, employs multi-layer neural networks to automatically learn hierarchical representations from large datasets, enabling complex applications like image recognition and natural language processing~\cite{LeCun2015DeepLearning}.}
\revi{The growing adoption of complex \ac{AI} models, particularly deep learning, has led to the prevalence of opaque ``black-box'' systems whose internal decision processes are difficult to interpret~\cite{Guidotti2019Survey}. \ac{XAI} addresses this challenge by providing methods to make model predictions more understandable to humans, thereby improving trust, accountability, and risk management in high-stakes automated decision-making contexts~\cite{Shin2021Effects, ferrario2022explainability}.} In \ac{XAI}, explanations can be classified as either \emph{global} or \emph{local}. \emph{Global} explanations offer insights into how a black-box model operates, aiding users in understanding its overall functioning \cite{Guidotti2019Survey}. In contrast, \emph{local} explanations focus on understanding the model's predictions for individual cases \cite{Guidotti2019Survey}. Additionally, the type of data processed by the \ac{AI} model also influences explanations. \ac{XAI} methods are most commonly applied to tabular, text, and image data, but they can also handle other types of data such as time series \cite{theissler2022Time}, audio \cite{akman2024audio}, and graphs \cite{Nandan2025GraphXAI}. A possible classification of \ac{XAI} techniques, based on the type of explanation used, is the following \cite{Guidotti2019Survey,Adadi2018Peeking}:

\begin{description}
    \item [Interpretable models] A small set of interpretable models,\revi{ such as \textit{decision trees} and \textit{decision rules},} can be used to explain more complex models by approximating their global or local behaviour. For example, a decision tree can approximate \revi{the behaviour of a} complex model, allowing its interpretation as a chain of \texttt{if-then-else} rules.
    \revi{This line of research builds upon foundational work in extracting symbolic rules from trained artificial neural networks \cite{Andrews1995RuleExtraction} and neuro-fuzzy systems \cite{Mitra2000NeuroFuzzy}, recently systematized in reviews on symbolic knowledge extraction and injection \cite{Ciatto24Symbolic}.}
    


    
    \item [Features Importance] Explanations may be provided by describing each feature's weight (i.e., importance) in the decision-making process. Such techniques are referred to as ``posthoc'' since they can be employed \emph{after} the complex model has been built. \revi{An example of such techniques is \ac{LIME} \cite{Ribeiro2016Why}, a technique to explain individual predictions by approximating the local behavior of a model with a simpler, interpretable model around a specific instance.}
    
    \item [Shapley Values] \revi{A particular instance of explanations based on feature importance is \ac{SHAP}, a technique based on cooperative game theory which uses \textit{Shapley values} to attribute a model's output to each input feature, ensuring consistency and local accuracy in explanations \cite{Lundberg2017Unified}. Compared to an approach like \ac{LIME}, which} can only provide local explanations, \ac{SHAP} can also produce global explanations, \revi{at the cost of a higher computational expense}. 
    
    \item [Saliency Masks] Saliency masks highlight subsets of the original instance (e.g., set of pixels of an image or words in a sentence) mainly responsible for a certain outcome. Thus, a ``salient mask'' visually highlights the aspects of the input instance determining the outcome. Saliency masks are generally used to explain deep neural networks. E.g., \ac{CAM} \cite{Zhou2016Learning}, and its generalisation \ac{Grad-CAM} \cite{Selvaraju2020GradCAM}, highlight the contents of an image based on an attention mechanism, which is visualised as an explanation. Saliency masks are also helpful for text: \emph{rationales} \cite{Lei2016Rationalizing} are a short and coherent piece of text (e.g., a sentence) that represents a sufficient subset of words that can be used to predict the original text on their own.
    
    \item [Sensitivity Analysis] This technique can be used to inspect black-box models by evaluating the correlation between the uncertainties of their inputs and outputs \cite{Saltelli2002Sensitivity}. Sensitivity analysis is generally used to develop visual tools for inspecting black boxes.
    
    \item [Partial Dependence Plot] \acp{PDP} are graphical representations showing the relationship between one or more input features and the model prediction while averaging the remaining features' effects. \acp{PDP} help understand how changes in certain input features affect the output of a black-box model. Individual Conditional Expectation \cite{Goldstein2015Peeking} is an extension of \acp{PDP} that allows for visualising the relationship between a feature and the prediction outcome, considering every individual instance in the dataset, instead of averaging over all instances.
    
    \item [\revi{Exemplar-based/}Prototype-based Explanations] A \emph{prototype} (or \emph{archetype}) is an object that is representative of similar instances (such as instances belonging to a certain class). Together with the outcome, these explanations return a prototype, which is very similar to the classified record, to highlight the rationale behind the prediction. Prototypes can be obtained, e.g., as a result of the averages of the features of a set of points \cite{Fouche2011Introduction}.

    \item [Counterfactual Explanations] Counterfactual explanations illustrate how minimal changes to input features can alter the model's prediction. They answer ``what-if'' questions by providing examples of how the input instance needs to be changed into a similar one to arrive at a different class in the output. \ac{DiCE} is a method that generates multiple diverse counterfactual instances for a given instance to help users understand different ways to achieve a desired outcome \cite{Mothilal2020Explaining}.
    
    \item [Neurons Activation] These interpretability techniques investigate how individual neurons (or groups of neurons) in neural networks respond to specific inputs. Among the most common neuron activation methods are \emph{activation maximisation} techniques \cite{Nguyen2016Synthesizing, Erhan2010Understanding}. These methods search for inputs that highly activate specific neurons or layers of a neural network model to unveil trained recognition patterns; this can produce a global interpretable model and can be particularly useful for image recognition models to identify which image patches are responsible for certain neural activations. 
    \item [\revi{Other}]\revi{ Other approaches include emerging or ad-hoc techniques that do not fall among the traditional categories. Examples of these are Explicit Factor Model for explainable recommendations \cite{Zhang2014ExplicitFactorModel}, Visual Interactive Model Explorer for \acs{ML} models debugging in sequential decision-making \cite{Antar24VIME}, approaches that exploit attention mechanisms for explaining deep neural networks \cite{Dehimi2024Attention}, and algorithmic-based \cite{desolda2023explanations} or agent-based explanations~\cite{Weitz2021Let}.}
\end{description}

\subsection{Towards Human-Centered Explainable Artificial Intelligence}\label{sec:hcxai}

Although XAI aims to \revi{address} the challenges posed by AI-based systems, which, despite their high precision, \revi{often} remain confined to research settings, it \revi{frequently} falls short of producing \emph{usable} systems that present their outcomes and explanations in a reliable, safe, and trustworthy manner. In real-world contexts, humans are generally unwilling or unable to trust \emph{black-box} systems that do not provide insight into their decision-making processes \cite{Combi2022Manifesto}. This lack of transparency prevents end-users from questioning \ac{AI} decisions, potentially allowing harmful biases to go unnoticed \cite{Stahl2023Unfair}. In response to these challenges, a new field of study has emerged in recent years at the intersection of \ac{HCI} and \ac{AI}, namely \ac{HCAI} \cite{Desolda2024Humancentered}. \ac{HCAI} proposes a new perspective on the interaction with \ac{AI}, aiming to augment rather than replace humans and their expertise \cite{Shneiderman2020HumanCentered}. \ac{HCAI} systems are designed to be ethically aligned, reflect human intelligence, and consider human factors \cite{Xu2019HumanCentered}. Recent research in \ac{HCAI} suggests the need for a truly human-centered process in the design of \ac{AI} systems, showing how there is no ``one-size-fits-all'' approach when designing \ac{AI} systems, as different user goals bring the need for a different level of automation and control in a system \cite{Esposito2024Fine}. One of the milestones of \ac{HCAI} is the adoption of \ac{HCI} methods for the design and development of \ac{HCAI}. In particular, \ac{HCAI} stresses the importance of user studies in eliciting requirements and validating final systems. \ac{HCD} is the general model adopted in \ac{HCI} for the design of systems that satisfy users' needs and expectations; it specifies that users are involved from the very beginning of the planning stage, and identifying user requirements becomes a crucial phase \cite{Rogers2023Interaction}. Various \ac{HCD}-based design processes are now available, such as \acs{UCD} Sprint \cite{Larusdottir2022Tutorial}. \ac{HCD} requires understanding who will use the system, where, and how to do what. The system is then designed by iterating a design-evaluation cycle. Being \revi{design-based} on empirical knowledge of user \revi{behavior}, needs, and expectations, it is possible to avoid serious mistakes and \revi{save reimplementation} time to correct such mistakes. In \ac{HCAI}, adopting \ac{HCD} is necessary when a direct interaction between users and the system is expected (e.g., in consumer applications, educational software, or health-related support). In this context, well-designed human-centred Human-\ac{AI} interfaces are essential for the success of an \ac{AI} system \cite{Shneiderman2022HumanCentered,Raees2024Explainable,Desolda2024Human,Esposito2024Fine}. Furthermore, when an \ac{AI} system is deployed in high-risk domains like medicine, explainability is crucial \cite{Gerdes2024Role,thePrecise4Qconsortium2020Explainabilitya,Combi2022Manifesto}. Therefore, a thoughtfully designed, human-centered \ac{XUI} is necessary.

\subsection{Explanation User Interfaces} \label{sec:xui} 
The notion of an interface mediating between the \ac{XAI} algorithm and the user was initially introduced by \ac{DARPA} \cite{Gunning2019DARPAs}.
\ac{DARPA} divides the \ac{XAI} process into two different stages: the generation of the explanations themselves, and the presentation of such explanations through \acp{UI}. The definition of this two-stage process is shared in the academic literature. \citeauthor{DBLP:conf/ijcnlp/DanilevskyQAKKS20} differentiate between explanation techniques and explanation visualizations: the former involves the generation of `rough' explanations (usually propounded by \ac{AI} researchers), while the latter concerns the ways `rough' explanations can be presented to users \cite{DBLP:conf/ijcnlp/DanilevskyQAKKS20}. A rigorous definition of \acp{XUI} is formulated by \citeauthor{Chromik2021HumanXAI}, who define a \ac{XUI} as  \emph{<<the sum of outputs of an XAI system that the user can directly interact with. An XUI may tap into the \ac{ML} model or may use one or more explanation-generating algorithms to provide relevant insights for a particular audience>>} \cite{Chromik2021HumanXAI}. In general, \acp{XUI} can be designed for various forms of explanation, both local and global, and can be presented in textual or visual form, or a combination of the two~ \cite{Bove2023Investigating}. \acp{XUI} can also leverage other modalities such as sound and tangible interfaces \cite{DBLP:conf/chi/SchneiderHRGTG21, DBLP:conf/mum/ColleyVH22, DBLP:conf/icmi/SchullerVRR0MD21}. 

According to the definition of \acp{XUI} \cite{Chromik2021HumanXAI}, two key aspects are fundamental to their development: \emph{interactivity} and \emph{adaptability} for different types of users.
\revi{Adaptive systems change their behavior automatically, driven by context-aware mechanisms, including models of their users and of specific tasks. Adaptive systems are important because there is no 'typical' user; there are many different users, and an individual user's requirements can change over time \cite{Fischer2023Adaptive}.}

\revi{Interactivity} plays a central role in the explanation process \cite{Miller2019Explanation}. Research in \ac{HCI} highlights the importance of designing interfaces that allow users to explore explanations freely \cite{DBLP:conf/chi/AbdulVWLK18}, ask follow-up questions, and access additional details as needed \cite{DBLP:conf/chi/KrausePN16}. Achieving this level of usability in real-world scenarios requires continuous user involvement throughout the engineering process, from the initial system design to the final testing phases \cite{Rogers2023Interaction}. However, despite its emphasis in \ac{HCI}, user-centred design remains relatively underexplored in \ac{XAI} and \acp{XUI}, highlighting a critical gap in the field.
\acp{XUI} fall into the category of interfaces designed for AI-infused systems. For these interfaces, traditional \ac{UI} principles remain essential; however, the unique and dynamic behaviour of AI systems necessitates the development of new guidelines tailored to their specific characteristics \cite{Amershi2019Guidelines}.

\subsection{Existing Literature Reviews} 

Many \acfp{SLR} and surveys have explored topics closely related to our \ac{SLR}, underscoring the growing relevance and urgency of advancing \ac{XAI} to ensure its practical implementation in real-world scenarios. These reviews often adopt distinct perspectives, focusing on different aspects of the topic. Several studies emphasise the concept of interactivity: \cite{Chromik2021HumanXAI} focuses on interaction design principles for \acp{XUI}, offering a structured analysis of user interactions and design guidelines. Similarly, \cite{Bertrand2023Selective} categorises interactive explanation techniques based on user evaluation \revi{constructs}, focusing on interactivity and its effects on user perception. \cite{Raees2024Explainable} further advances the argument by examining interactivity in \ac{XAI} systems through the lens of Human-Centred AI literature, proposing the evolution of \ac{XAI} into Interactive \ac{AI}. Another study \cite{DBLP:conf/interact/MuralidharBOA23} examines the components of transparency in \ac{AI} systems, aiming to enhance user interpretability and proposing strategies for designing systems that are more interpretable and explainable to users. The evaluation of \ac{XAI} applications through user studies represents another important area of focus. \cite{RongLNFQUSKK24} investigates how these studies are conducted, while \cite{AlAnsari2024UserCenteredEO} analyses the role of \ac{HCI} techniques in improving \ac{XAI} goals.
\ac{HCI} remains central to the exploration of explanation design in
\cite{Ferreira2020What}, whose main purpose is to investigate who the recipients of \ac{AI} explanations are, the motivations for providing them, and the methods for aligning explanations with user needs.
The intersection between \ac{XAI} and \ac{HCI} is also investigated in \cite{SystematicsXAIforAI-HCI}, where the authors highlight the opportunities for \ac{HCI} concerning \ac{XAI} tools.
Furthermore, studies like those by \cite{Laato2022How} and \cite{DBLP:journals/corr/aifromuserperspective}focus on how explanations can be effectively tailored to specific user categories, particularly non-technical and end users, addressing critical gaps in accessibility and usability.
Finally, \cite{DBLP:conf/vl/FilhoBO24} narrows its scope to reliance-aware explainable user interfaces, synthesising secondary studies to highlight design solutions to foster appropriate reliance in \ac{XAI}-assisted decision-making.

\paragraph{\revi{Novelty of our SLR}}
\revi{Unlike other research}, our review examines the literature on \ac{XUI}, leveraging both algorithmic foundations and \revi{human-centered} design principles.
Specifically, \revi{the current} research investigates three aspects:
\begin{enumerate*}[label=\roman*.]
    \item the contextual elements that guide the design of \ac{XUI} solutions;
    \item \ac{AI} and \ac{XAI} methodologies commonly employed in the development of \acp{XUI};
    \item the evaluation methods, driven by \ac{HCI} principles, used to assess \acp{XUI} effectively.
\end{enumerate*}

Moreover, we also discuss the design principles that emerge from the literature.
Unlike prior works that often treat these aspects in isolation, our review seeks to integrate these dimensions to identify design trends among human-centred \acp{XUI} and \ac{XAI} algorithms by synthesising insights from 146 papers retrieved from IEEE, ACM, and Scopus.
Our contribution lies in providing a comprehensive overview that progresses the discussion on \acp{XUI} by connecting the technical, design, and evaluation components in a unified way, addressing the need for integrated solutions in real-world \ac{XAI} applications.

\section{Planning and Conducting the Systematic Literature Review} \label{sec:methods}  
We used a rigorous, reproducible methodology to perform an \ac{SLR}. According to \citeauthor{Kitchenham2004Procedures}, conducting an \ac{SLR} involves three main stages: planning, execution, and reporting \cite{Kitchenham2004Procedures}. This section details the first two stages, while the third one is covered in \Cref{sec:RQ1,sec:RQ2,sec:RQ3,sec:RQ4}.

\subsection{Planning the SLR}
Planning an \ac{SLR} is a multi-step process that involves the following activities \cite{Kitchenham2004Procedures}:
\begin{enumerate*}
    \item formulating a research question; 
    \item defining the set of search strings; 
    \item electing the data sources; 
    \item defining the inclusion and exclusion criteria. 
\end{enumerate*}
In this section, we report the details of each activity, discussing the motivations behind every choice.

\subsubsection{Formulation of the Research \revi{Questions}}\label{sec:research-questions}

The main goal of our \ac{SLR} is to assess the current state of research about \aclp{XUI} and to understand how AI models, tasks, different domains, and users affect the way \acp{XUI} are designed and evaluated. With this objective in mind, we formulated the following research questions:

\begin{enumerate}[label=(RQ\arabic*),align=left,leftmargin=1.25cm]
    \item What influences the design of \acp{XUI}?
    \item What \ac{XAI} solutions \revi{(e.g., tools, frameworks, AI models)} are used to develop \acp{XUI}?
    \item What solutions are used to evaluate \acp{XUI}?
    \item What can guide the design of \acp{XUI}?
\end{enumerate}

\subsubsection{Definition of the Search Strings}\label{sec:search-strings}

We formulated our search strings by deriving keywords from two main sources: \begin{enumerate*}[label=(\roman*)]
\item an initial assessment of some of the most highly cited papers on \acp{XUI}, and
\item the authors' expertise in the subject matter.
\end{enumerate*}
The search strings were designed around three primary concepts aligned with our research questions. Each concept, along with its synonyms and variations, was included to ensure comprehensive coverage of the literature.

\begin{description}
\item \emph{\acf{XAI}}: This encompasses all techniques that provide insights into the inner workings of black-box models or involve the design of inherently interpretable (white-box) models \cite{Guidotti2019Survey}. To capture relevant studies, we included variations of terms such as ``\acs{XAI}'' and ``\acl{XAI}''. The stem ``expla*'' was employed to include the latter and also terms like ``explainability'', ``explanations'', and related variations.
\item \emph{Human-Centered Design}: High-quality user interfaces require a human-centered approach. To capture this aspect, we made sure to include keywords that could encompass terms such as ``human-centred'', ``human-centered'', ``user-centred'', and ``user-centered'', accommodating both British and American spellings. The broader terms ``human'' and ``user'' were thus used to ensure inclusivity.
\item \emph{Interactive User Interfaces}: Since \acp{XUI} inherently involve user interfaces, relevant literature must address this aspect. Keywords like ``interface'' and the stem ``interact*'' were included to cover terms such as ``interactivity'', ``interaction'', and ``interactive''.
\end{description}

The resulting final search string was defined as follows:
\begin{quote}
("xai" OR "expla*")
AND ("user" OR "human")
AND ("interface" OR "interact*")
\end{quote}
This formulation ensures a comprehensive yet precise retrieval of relevant studies, aligning with the scope of our \ac{SLR}.
\revi{In particular, our goal was to identify papers that simultaneously address all three areas — eXplainable Artificial Intelligence, Human-Centered Design, and Interactive User Interfaces — rather than focusing on each aspect in isolation. These three dimensions are further operationalized and analyzed according to the contribution types described in \cref{sec:datasyntesis}.}

\subsubsection{Selection of Data Sources}
We started this \ac{SLR} from scientific digital libraries. The examined libraries were ACM Digital Library\footnote{\url{https://dl.acm.org}}, IEEE Xplore\footnote{\url{https://ieeexplore.ieee.org}}, 
and Scopus\footnote{\url{https://www.scopus.com/}}. During the definition of the search string, we noticed that the same search has to be performed differently depending on the library (i.e., using a different syntax). Each library contained various options for searching content: for example, they allow searching for keywords within the article title, abstract, full text, or all of the above. We searched for the most comprehensive choice available for each digital library. All searches were conducted on the entire database due to the inherent multidisciplinarity of \acp{XUI} applications. \revi{Google Scholar was omitted, as it does not meet the inclusion criteria based on peer-reviewed publications and venue quality. Moreover, since the selected databases already cover most high-quality venues indexed by Google Scholar, its inclusion would mainly have increased the number of duplicates without improving coverage.}

\subsubsection{Definition of the Inclusion and Exclusion Criteria}

\begin{table}[t]
\centering
\small
\caption{Inclusion and Exclusion Criteria}
\label{tab:inc_ex_criteria}
\begin{tabularx}{\textwidth}{@{} lXX @{}}
\toprule
Criteria & Inclusion Criteria & Exclusion Criteria \\
\midrule
Date & Published in or after 2013 & Published before 2013 \\
Language & Written in English & Not written in English \\
Type of Publication & Full and short papers & Book chapters, extended abstracts, workshop proposals, posters, and demo papers \\
Peer Review & Published in A*, A, or B conferences, or Q1/Q2 journals; C-ranked conferences or Q3 journals are carefully evaluated & Published in Q4 journals, unranked conferences, or national venues \\
\bottomrule
\end{tabularx}
\end{table}

This step concerns the final selection of the relevant publications.
\citeauthor{Kitchenham2004Procedures} distinguishes between practical inclusion and exclusion criteria, which address logistical constraints, and quality-based criteria, which are derived from the research question to refine the selection of relevant studies. \Cref{tab:inc_ex_criteria} outlines the practical criteria applied in this study. Notably, we focused on publications starting from 2013 since \ac{XAI} was formalized only in 2017 \cite{Gunning2019DARPAs}, but concerns on biases and the need for explanations started earlier, as discussed in \cite{Minsky2013Society}. Additionally, inclusion criteria were established to ensure the selected papers aligned with the research objectives. These criteria focus on the content of the research question and require that a study:

\begin{enumerate*}[label=(\roman*)]
    \item Involves a user study as part of its evaluation;
    \item Incorporates an explanation algorithm;
    \item Is based on an AI model, or, if the AI system is not explicitly defined, is designed with AI in mind (e.g., Wizard of Oz studies).
\end{enumerate*}

This approach ensured that only studies contributing to the advancement \ac{XAI} and \ac{XUI} were considered in the analysis.


\subsection{Conducting the Literature Review}
After planning, the literature review was conducted. As Kitchenham suggests, the review consisted of two main activities \cite{Kitchenham2004Procedures}: the literature review execution and the data synthesis. The subsequent subsections detail both steps.

\begin{figure}[t]
    \centering
    \includegraphics[width=0.7\linewidth]{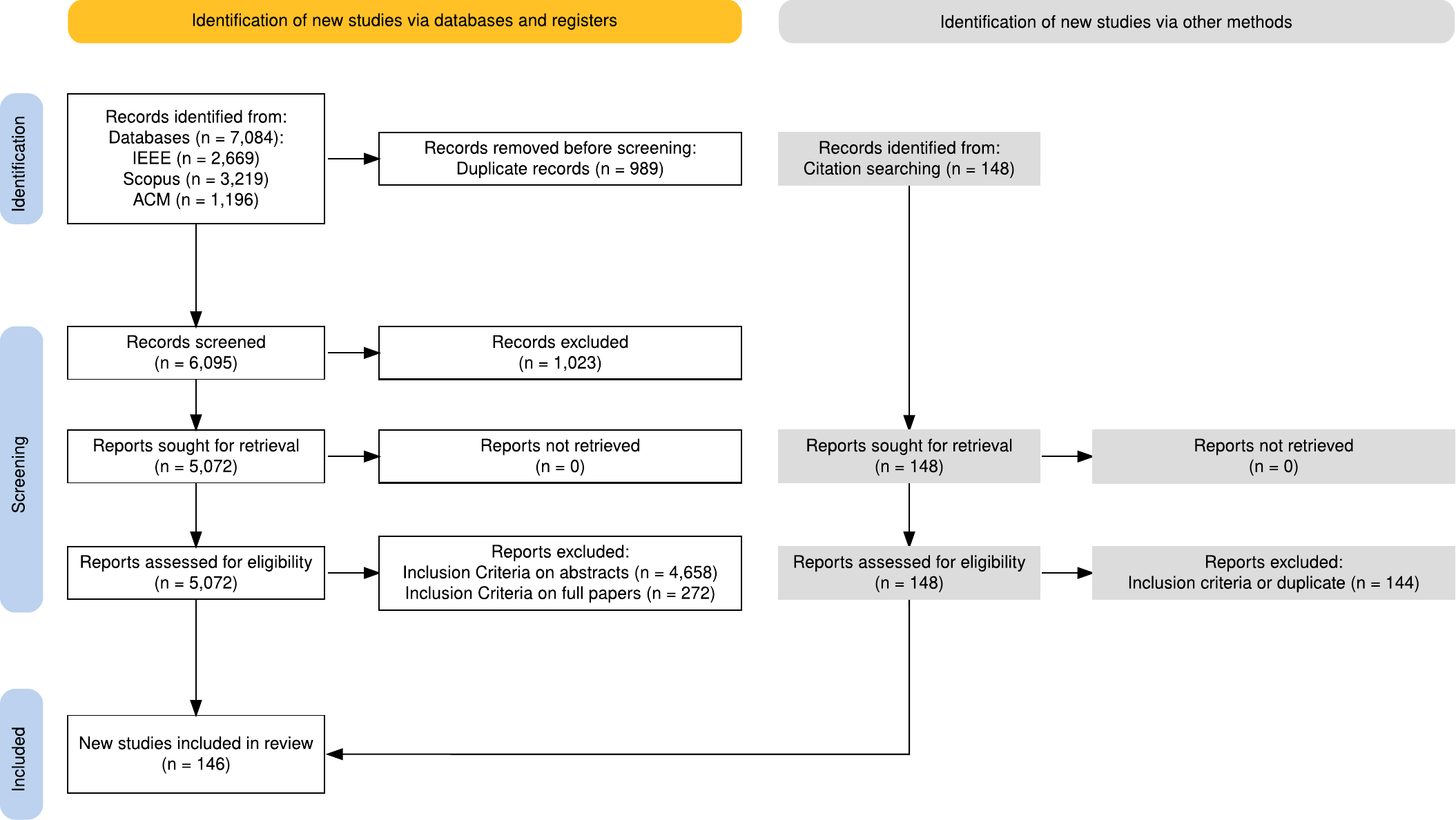}
    \caption{PRISMA \cite{Page2021PRISMA} flow diagram depicting the identification, screening, eligibility, and inclusion process of studies in the systematic literature review.}
    \label{fig:prisma}
    \Description{PRISMA flow diagram depicting the identification, screening, eligibility, and inclusion process of studies in the systematic literature review.}
\end{figure}

\subsubsection{Literature Review Execution}
The activity was performed from June 2024 to December 2024, following the process depicted in \cref{fig:prisma}, which mainly consists of two phases:
\begin{enumerate}
    \item \emph{Phase 1 -- Digital Library Search}: We searched the identified sources using the previously described search strings.
    \item \emph{Phase 2 -- Backward and Forward Snowballing Search}: We checked references and citations of the publications resulting from the previous phase and publications that cited publications from Phase 1 \cite{Wohlin2014Guidelines}.
\end{enumerate}

The initial search across the digital library yielded a total of 7084 potentially relevant publications. After a duplicate check, a dataset of 6095 publications was obtained. After filtering, using the previously defined inclusion criteria, 1023 publications were excluded as they did not meet the quality criteria, leaving a final dataset of 5072 candidates. After this first filtering step, the title and abstract were analyzed, resulting in the exclusion of a further 4658 publications, reducing the number to 414. Filtering was conducted independently by three researchers; \revi{any conflict was reported and addressed in} the next step.
\revi{Disagreements were resolved through discussion among the researchers, and in cases where one researcher disagreed with the other two, a majority vote was used. Inter-rater agreement on inclusion/exclusion decisions was considerable, with a Cohen’s $\kappa$ of $.81$. Through successive discussions, full inter-rater agreement was reached.}
The publications were then carefully analyzed by accessing their full text, narrowing down to 142 valid publications for the review. After forward and backward snowballing, an additional 148 candidates were retrieved and filtered down to 4 using the same inclusion criteria. In conclusion, the final dataset comprises a total of 146 publications.

\subsubsection{Data Synthesis}\label{sec:datasyntesis}

We classified the types of contributions into three distinct categories:
\begin{description}
    \item Type 1 - UI: Publications that focus on presenting the design of an \ac{XUI} tailored for a specific application.
    \item Type 2 - Framework: Methodological papers that introduce design guidelines for \acp{XUI} or propose frameworks to support the explanation process.
    \item Type 3 - Both: Papers that combine both aspects, presenting an \ac{XUI} and proposing a set of design guidelines or a framework to support its development or evaluation.
\end{description}
For each paper, multiple categorical dimensions were collected for analysis, which align with established taxonomies to facilitate a structured evaluation of the results. \revi{Furthermore, for papers that reported insights for designing \ac{XUI} systems---such as principles, recommendations, lessons learned, etc.---we collected design guidelines and frameworks.} The knowledge extracted from the selected literature is reported in \Cref{sec:results}.

\section{Results}\label{sec:results}

\revi{The 13 dimensions identified in the literature review predominantly reflect either a human-centered or algorithmic perspective. For example, aspects related to user validation (e.g., visual mode, type of study) belong to the former category, while aspects related to the AI model (e.g., AI task, type of data) fall into the latter. The only exception is the \textit{evaluation construct} dimension, which is equally applicable to both perspectives. Moreover, data from each dimension were used to answer one or more research questions.} A complete description of the \revi{extracted} dimensions and their corresponding research questions is presented in \cref{tab:research_dimensions}. 

\begin{table}[t]
\centering
\small
\caption{Overview of research dimensions, associated research questions, and their descriptions.}
\label{tab:research_dimensions}
\renewcommand{\arraystretch}{1.3}
\begin{tabularx}{\textwidth}{@{} lccX @{}}
\toprule
Dimension & \revi{Perspective} & Research Questions & Description \\ \midrule
Application domain & \revi{Human-centered} & RQ1 & Refers to the specific field or sector where the Explainable AI (XAI) system is applied, such as healthcare, finance, or education. \\
Type of data & \revi{Algorithmic} & RQ1 & Indicates the nature of the data used by the AI system, such as tabular, image, text, or audio. \\
AI & \revi{Algorithmic} & RQ2 & The specific AI model or architecture employed, such as neural networks, decision trees, or ensemble methods. \\
AI Task & \revi{Algorithmic} & RQ2 & Describes the goal of the AI system, such as classification, regression, clustering, or prediction. \\
Output Type & \revi{Algorithmic} & RQ2 & Refers to the format of the AI model's outputs, such as probabilities, labels, and rankings. \\
XAI Techniques & \revi{Algorithmic} & RQ2 & The methods used to generate explanations, such as counterfactual reasoning, feature importance, and Shapley values. \\
Explanation Modality & \revi{Human-centered} & RQ2 & The mode of explanation delivery, such as textual, visual, natural language, or a combination of these. \\
Interactivity & \revi{Human-centered} & RQ2 & Indicates whether the XUI system implements interactive elements to enhance user engagement and understanding. \\
Visual Mode & \revi{Human-centered} & RQ2 & The type of visual representation used in the interface, such as heatmaps, trendlines, or other visual models. \\
Type of Study & \revi{Human-centered} & RQ3 & The methodological approach used to evaluate the Explanation User Interface, such as interviews, controlled experiments, or user observations. \\
Type of Users & \revi{Human-centered} & RQ1, RQ3 & The target audience for the system, such as domain experts, end-users, or AI specialists. \\
Number of Participants & \revi{Human-centered} & RQ3 & The sample size of users involved in studies evaluating the system. \\
Evaluation constructs & \revi{Both} & RQ3 & The criteria used to assess the effectiveness of the XAI system, such as trust, usability, understanding, or task performance. \\
\bottomrule
\end{tabularx}
\end{table}

\subsection{\texorpdfstring{RQ1. What influences the design of \acp{XUI}?}{RQ1. What influences the design of XUIs?}}\label{sec:RQ1}

This research question investigates \revi{the factors that influence} \acp{XUI} design, identifying the critical elements (e.g., constraints and considerations) that guide it. Our analysis examines the problem space from two perspectives. From a human-centred perspective, we focus on the domain of the application and the target users to whom the interface is designed. From an algorithmic perspective, we analyse \revi{the type of data utilised by} the underlying AI layer.
\paragraph{\revi{Summary of findings}} \revi{High-stakes domains (e.g., health, finance) more commonly employ \acp{XUI}, which are generally used by domain experts (e.g., physicians). Moreover, \ac{AI} systems behind \acp{XUI} typically work with tabular and image data (e.g., patient data, medical scans). \acp{XUI} used in the finance domain, albeit high-stakes, are also intended for non-expert users and tend to utilize time-series and tabular data.}

\begin{table*}[t] 
    \centering
    \small 
    
    \begin{minipage}[t]{0.30\textwidth}
        \centering
        \caption{User Types Dist.}
        \label{tab:user_types}
        \begin{tabular}{@{} lc @{}}
            \toprule
            Type of Users & Count \\
            \midrule
            Domain-experts & 48 \\
            Non-experts & 38 \\
            AI experts & 22 \\
            Not specified & 21 \\
            \bottomrule
        \end{tabular}
    \end{minipage}%
    \hfill 
    \begin{minipage}[t]{0.36\textwidth} 
        \centering
        \caption{App. Domains Dist.}
        \label{tab:domain_data}
        \begin{tabular}{@{} lc @{}}
            \toprule
            Application Domain & Count \\
            \midrule
            Health & 38 \\
            Finance/Economics & 20 \\
            General & 15 \\
            NLP & 13 \\
            AI \& Robotics & 12 \\
            Media \& Comm. & 10 \\
            Education & 9 \\
            Rec. Systems & 7 \\
            Network & 5 \\
            Mobility & 4 \\
            Cybersecurity & 3 \\
            Other & 3 \\
            Agriculture & 2 \\
            \bottomrule
        \end{tabular}
    \end{minipage}%
    \hfill 
    \begin{minipage}[t]{0.30\textwidth}
        \centering
        \caption{Data Types Dist.}
        \label{tab:data_types}
        \begin{tabular}{@{} lc @{}}
            \toprule
            Type of Data & Count \\
            \midrule
            Tabular Data & 61 \\
            Images & 26 \\
            Text & 19 \\
            Time Series & 13 \\
            Video & 6 \\
            Multimodal & 4 \\
            Audio & 2 \\
            Unknown & 4 \\
            Generic & 3 \\
            \bottomrule
        \end{tabular}
    \end{minipage}
    
\end{table*}
\subsubsection{\revi{Application Domains}}
The analysis of application domains, reported in \Cref{tab:domain_data}, highlights the prominence of \acp{XUI} in high-stakes fields such as health and finance \cite{He2024VMS, Esfahani2024Preference, Barda2020Qualitative, Wang2021CNN, DBLP:journals/ijmms/KimCPPNJL23, DBLP:conf/chi/BhattacharyaEXMOS2024,MezaMartinez2023Does, Purificato2023Use}. These domains, characterised by their critical decision-making requirements, align closely with the ones in which \ac{XAI} is extremely useful \cite{Combi2022Manifesto}. However, underexplored areas like mobility \cite{Antar24VIME, DBLP:conf/chi/SchneiderHRGTG21}, cybersecurity \cite{desolda2023explanations, stites2021sage, wu2021android}, and education \cite{DBLP:conf/cscw/ShenHWH23, Shin2022XDesign} suggest opportunities for expanding the impact of \acp{XUI} into new contexts. The number of domains reinforces the broad applicability of \acp{XUI} solutions across various sectors.

\subsubsection{\revi{Type of Users}}

\Cref{tab:user_types} illustrates the distribution of user types targeted by \ac{XUI} designs. The results reveal a predominant focus on domain experts, reflecting the necessity for explainability tools in decision-critical tasks within specialised industries such as healthcare and finance \cite{Wysocki2023Assessing,Bhattacharya2023Directive,Ferrario2020ALEEDSA}. Non-experts constitute the second-largest group, highlighting efforts to democratise AI and make it accessible to broader audiences \cite{Wang2021CNN,Bove2022Contextualization}. In contrast, interfaces targeting \ac{AI} experts are less common \cite{Shin2022XDesign, Antar24VIME}, possibly due to their familiarity with \ac{AI} systems. The substantial proportion of studies labeled as ``Not specified'' emphasises the need for clearer identification of user personas in future research. Furthermore, user studies often utilised platforms such as Prolific\footnote{\url{https://www.prolific.com}} and Amazon Mechanical Turk\footnote{\url{https://www.mturk.com}}, indicating a reliance on online participant recruitment for evaluation purposes \cite{Dominguez2020Algorithmic, Guo2022Building, Alqaraawi2020Evaluating, Mucha2021Interfaces, Guesmi2022Explaining}.

\subsubsection{\revi{Data Types}}
The distribution of data types in \ac{XUI} designs is presented in \revi{  \cref{tab:data_types}}. Tabular and image data prevail, reflecting the prevalence of structured datasets and the importance of images in the most-explored domain (i.e., medicine) \cite{Hwang2022Clinical, Gromowski2020Process}. This also highlights the importance of visual representation in explainability. Other types, like text \cite{Riveiro2021Thats} and time series \cite{Upasane2021Big}, indicate a growing interest in different data modalities. However, the relatively smaller focus on multimodal and media-based data \cite{Wang2022M2Lens} reveals potential for future exploration, particularly in complex scenarios requiring the integration of multiple data types.

\subsubsection{\revi{Interplay Among Application Domain, User Type, and Data Type}}
\begin{figure}[t]
  \centering
  \includegraphics[width=0.60\textwidth]{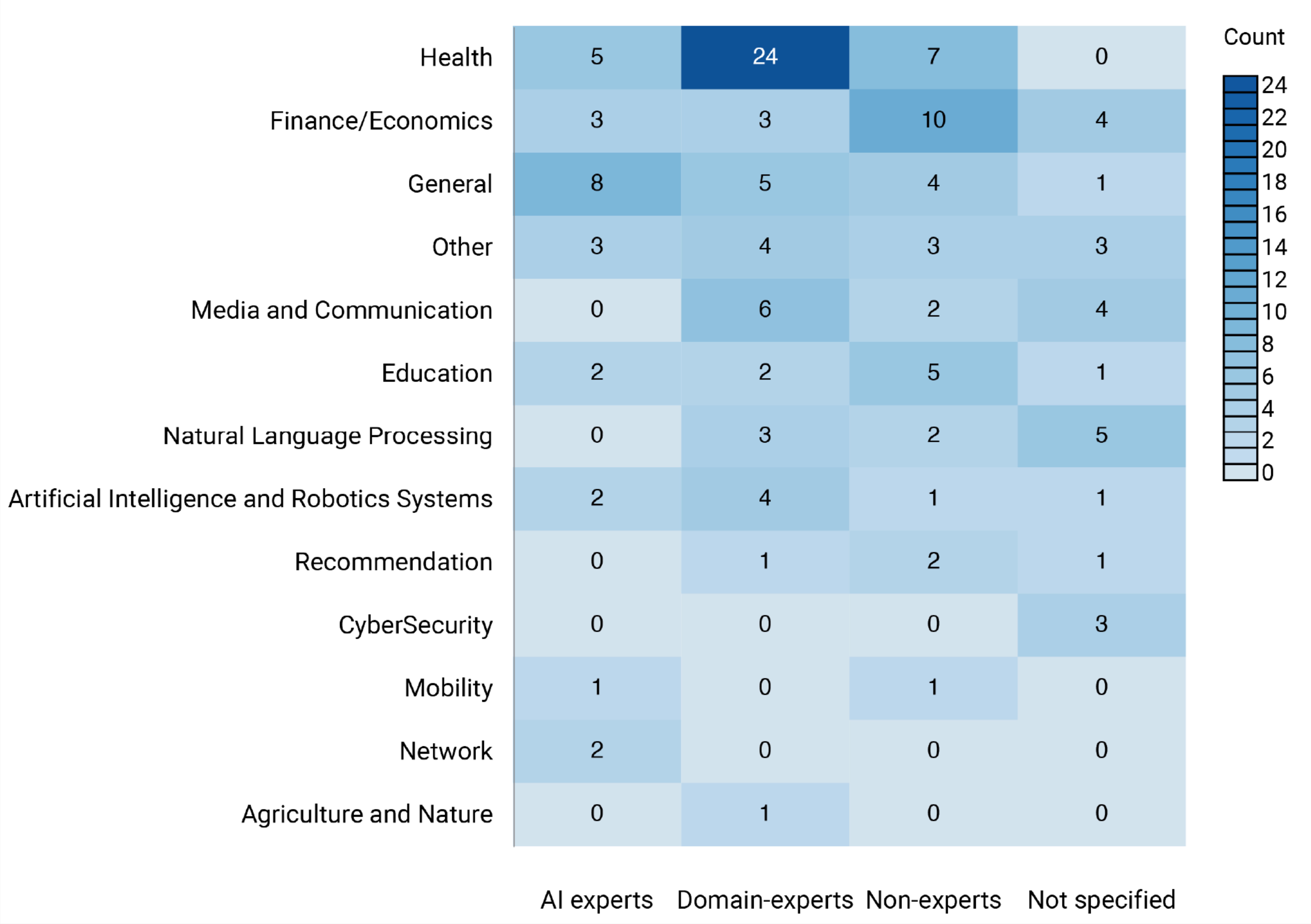}
  \caption{Heatmap showing the relationship between user types and application domains, illustrating how different user categories are distributed across various domains}
  \label{fig:heatmap_domain_user}
  \Description{\revi{ Relationship between user types and application domains. The figure illustrates how different user categories are distributed across various domains}}
\end{figure}

Further insights from \Cref{fig:heatmap_domain_user} confirm earlier observations about the strong relation between user types and application domains. The health domain emerges as strongly associated with domain experts and both tabular data and images, reflecting the importance of precise and trustworthy explanations in high-stakes contexts such as medical diagnostics and healthcare decision-making \cite{Combi2022Manifesto}. Here, \acp{XUI} are often designed to align with the expertise of professionals, ensuring that these tools support accuracy, trust, and informed decisions \cite{DBLP:journals/pacmhci/Okoloeasy24, DBLP:conf/chi/Zhang24Rethinking, Naiseh2023How, Lamy2019Explainable}.

Similarly, the finance and economics domain demonstrates a dual focus, catering to both domain experts \cite{Purificato2023Use, DBLP:conf/caise/FusslNH24}, such as financial analysts, and non-experts, such as general users of financial services \cite{Bove2022Contextualization, Chromik2021Making, DBLP:conf/fat/BertrandEM23}. Most of the time, \ac{XUI} systems in this domain focus on time series and tabular data. This reflects the need for \acp{XUI} that provides clear, interpretable insights for specialists while ensuring accessibility and usability for broader audiences. \revi{The media} and communication domain displays a balanced representation of general users and domain experts (without a strong focus on a specific data type) \cite{Anderson2020Mental, Wu2022AIDriven, Khanna2022Finding}, underscoring the necessity of designing interfaces that make AI-driven explanations accessible to diverse audiences, including professionals and general users. However, \ac{AI} and robotic systems are predominantly associated with \ac{AI} and domain experts \cite{Piorkowski23AIMEE, Baniecki2024Grammar, Nourani2022DETOXER, Mishra2022Why}. This trend aligns with the technical expertise required to develop, understand, and evaluate such systems. For general-purpose applications, the findings indicate a noticeable emphasis on AI experts, suggesting that such interfaces often prioritise technical depth and flexibility. These systems are frequently foundational tools or frameworks designed to be adaptable across various domains, appealing to users with advanced AI knowledge who can tailor them to specific use cases.

\subsection{\texorpdfstring{RQ2. What XAI solutions (e.g., tools, frameworks, AI models) are used to develop \acp{XUI}?}{RQ2. What XAI solutions (e.g., tools, frameworks, AI models) are used to develop XUIs?}}\label{sec:RQ2}

\begin{figure}
    \centering
    \includegraphics[width=0.8\linewidth]{img/heatmap_ai_xai_t_rebuttal.png}
    \caption{Heatmap showing the comparison between \ac{AI} models and \ac{XAI} techniques, illustrating the most frequent usage of XAI techniques with AI models identified in the analyzed literature.}
    \label{fig:heatmapAIXAI}
\end{figure}

This research question \revi{has a more algorithmic-centric perspective and} examines the connection between the algorithms used to generate explanations and how these are conveyed through \acp{XUI}. To structure the analysis, we utilised the taxonomy proposed by \citet{Guidotti2019Survey}, as outlined in \Cref{sec:xai}, which categorises \ac{XAI} techniques based on their methodological characteristics and their application to \ac{AI} models. Additionally, we consider the framework described by \citet{Gunning2019DARPAs} \revi{to classify explanation} modalities (i.e., how explanations are conveyed to users through different \revi{media}).
\revi{In this context, the term solution is intentionally used in a broad sense, referring to any \ac{XUI} component or approach---consistent with the \ac{DARPA} \ac{XAI} definition---that mediates the interaction between the user and the explanation provided by an \ac{XAI} system (See \Cref{sec:xui}).}

\paragraph{\revi{Summary of findings}} \revi{The most common \ac{XAI} solutions in \acp{XUI} are feature importance, counterfactual explanations, and Shapley values, while the most common \ac{AI} models are neural networks, ensembles, and transparent models. Some \ac{XAI} solutions are more present for certain \ac{AI} models (e.g., neural activation techniques are used with neural networks, explanation decision trees are used with reinforcement learning models). Most \acp{XUI} use visual-based explanations, most commonly employing heatmaps, barcharts, and trend lines. Textual explanations are also widely used. \ac{XUI} systems often consist of interactive \ac{UI}.}

\subsubsection{\revi{Relationships Between AI Systems and Explainability Techniques}}\label{sec:rq2-ai&xai}

\Cref{tab:AI_alg} summarises the \ac{AI} algorithms employed in the surveyed studies. Neural networks prevail over the other techniques, followed by ensemble methods and white-box models. These results highlight the prevalence of complex, opaque models like neural networks, which require \ac{XAI} techniques to enhance interpretability. The distribution of XAI methodologies, as shown in \Cref{tab:XAI_method}, reveals that feature importance methods are the most frequently used, followed by counterfactual explanations and Shapley values. These techniques primarily aim to provide localised insights into individual model predictions. The dominance of neural networks and the frequent use of these explanation techniques underscore the need for some form of interpretability in such complex systems. Feature importance, counterfactual explanations, and Shapley values are valuable tools to bridge the gap between black-box models and actionable, user-friendly explanations. As emphasised by Miller, effective explanations should focus on relevance and comprehensibility rather than exhaustively detailing model structure \cite{Miller2019Explanation}. It should still be noted that, although the literature employs such techniques to provide some form of explanation for the black-box model, this has a polarising effect as some argue that this is, in fact, not the best course of action, instead preferring white-box models when possible \cite{Rudin2019Stop}.

\begin{table}[t]
\centering
\begin{minipage}{0.45\textwidth}
    \centering
    \small
    \caption{AI Algorithm}
    \label{tab:AI_alg}
    \begin{tabular}{@{}cc@{}}
        \toprule
        AI algorithm & Count \\ \midrule
        Neural Network & 51 \\
        Ensemble &  17\\
        White Box & 12 \\
        Probabilistic & 7 \\
        Math & 5 \\
        Rule-based & 5 \\
        Reinforcement Learning & 5 \\
        Agnostic & 3 \\
        Other Black-box & 4 \\
        Fuzzy & 1 \\
        Unknown & 13 \\ \bottomrule
    \end{tabular}
\end{minipage}%
\hfill
\begin{minipage}{0.45\textwidth}
    \centering
    \small
    \caption{Explainable AI Techniques}
    \label{tab:XAI_method}
    \begin{tabular}{@{}cc@{}}
        \toprule
        Explanation Techniques & Count \\ \midrule
        Features Importance & 32 \\
        \revi{Counterfactual}/Exemplars &  28\\
        Shapley Values & 27 \\
        Exemplars/\revi{Prototype-based} & 19 \\
        Decision Rules & 14 \\
        \revi{Saliency} Mask & 12 \\
        Neurons Activation & 7 \\
        Decision Trees & 4 \\
        Partial \revi{Dependence} Plot & 4 \\
        Sensitivity Analysis & 1\\
        \textbf{\revi{Other}} & \textbf{12} \\
        \bottomrule
    \end{tabular}
    \end{minipage}
\end{table}
Some less common techniques, such as salient masks \cite{Hwang2022Clinical, Kadir2023User}, neuron activation \cite{Wang2023DeepSeer}, decision trees \cite{Khanna2022Finding}, and partial dependency plots, suggest specialised applications in \acp{XUI}. These methods often target unique requirements or particular user groups, emphasizing the diverse needs within the \acp{XUI} design space.

\revi{Regarding} the relationship between \ac{XAI} techniques and \ac{AI} algorithms, several recurring patterns can be observed. Feature importance is prominently linked to neural networks \cite{Lee2020CoDesign, Hernandez-Bocanegra2021Effects, DBLP:conf/aime/RohrlMLKHKSHD23, Lee2023LIMEADE, Heimerl2022Unraveling, Mishra2022Why, Shin2022XDesign}, ensembles \cite{Malandri2023ConvXAI, DBLP:conf/chi/BhattacharyaEXMOS2024,He2024VMS}, and probabilistic methods \cite{Riveiro2021Thats}, reflecting its adaptability across diverse AI models. Counterfactual explanations are broadly applied, particularly in ensemble \cite{Malandri2023ConvXAI, Ma2022Explainable, Bove2023Investigating, Cau2023Supporting} and neural network settings \cite{Sovrano2021Philosophy, Esfahani2024Preference, DBLP:conf/chi/Zhang24Rethinking, Hao23Time}. Shapley values are closely associated with neural networks \cite{DBLP:journals/tlt/LuWCZ24, kim2023designing, MezaMartinez2023Does, Wang2022M2Lens,He2024VMS} and probabilistic models \cite{Malandri2023ConvXAI, Purificato2023Use, Cheng2022VBridge}, leveraging their utility in quantifying feature contributions in complex systems. Techniques like neuron activation are nearly exclusive to neural networks \cite{Hwang2022Clinical, Wang2023DeepSeer, Nourani2020Investigating, DBLP:conf/cikm/AhnHHK24, wu2021android} due to their specific relevance to such architectures. Decision rules appear with rule-based \cite{Piorkowski23AIMEE,Sevastjanova2021QuestionComb}, ensemble \cite{DBLP:conf/chi/BhattacharyaEXMOS2024, Wang2019Designing} and Neural Network \cite{Ming2019RuleMatrix, MezaMartinez2023Does,Krause2017Workflow}, while decision trees are used along with reinforcement learning techniques \cite{Khanna2022Finding,Kridalukmana2022SelfExplaining}.

Overall, the choice of \ac{XAI} techniques often depends on the type of \ac{AI} algorithm being explained. \revi{While some \ac{XAI} techniques are specifically designed for particular model architectures}, such as neural activation analysis for neural networks \cite{Wang2023DeepSeer,Hwang2022Clinical,Nourani2020Investigating}, \revi{or analyzing decision trees in reinforcement learning \cite{Kridalukmana2022SelfExplaining,Khanna2022Finding}. Rule extraction techniques, although associated with rule-based models \cite{Piorkowski23AIMEE,Sevastjanova2021QuestionComb}, can be effectively applied to a wide variety of predictive models to generate interpretable insights, regardless of the underlying algorithm \cite{Sabbatini25Survey, Gromowski2020Process,MezaMartinez2023Does,Metta2021Exemplars,Ming2019RuleMatrix}. Model-agnostic methods, on the other hand,} include feature importance, Shapley values, and counterfactual examples. These models are widely adopted due to their adaptability across various contexts and model architectures \cite{stites2021sage,Riveiro2021Thats,Malandri2023ConvXAI,DBLP:journals/pacmhci/Okoloeasy24}.

\subsubsection{\revi{Explanation Modalities}}\label{sec:rq2-explanation-modality}

\begin{figure}[t]
    \centering
\includegraphics[width=0.35\textwidth]{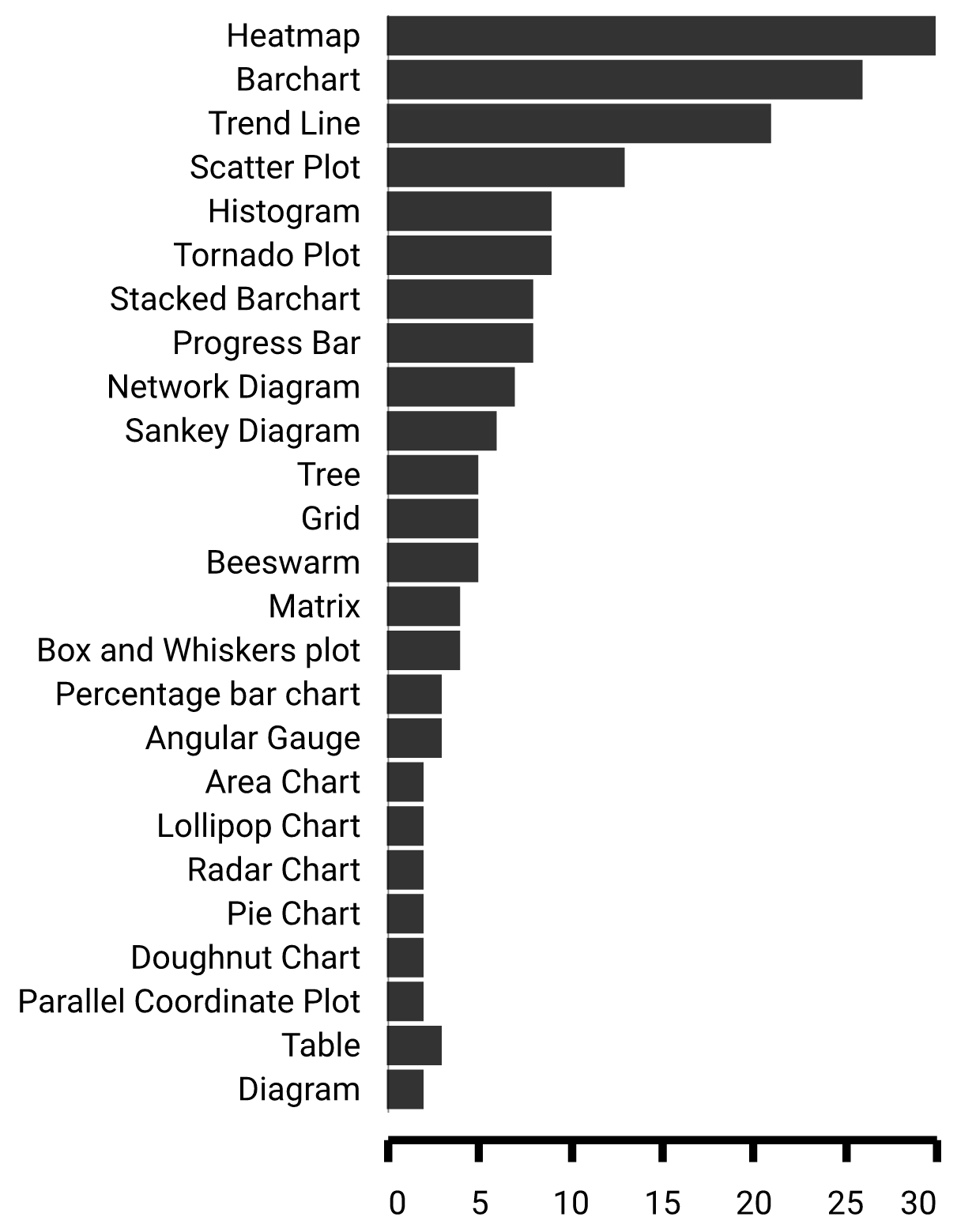}
    \caption{Frequency of visualisation techniques used in explanation user interfaces. Heatmaps, bar charts, and trend lines are the most frequently used. A wide range of other formats, such as scatter plots, histograms, Sankey diagrams, and tree structures, are also represented, reflecting the diversity of visual explanation strategies.}
    \label{fig:visual_tax_bar}
    \Description{Frequency of visualisation techniques used in XUIs. The bar chart presents the types of visualisations employed to convey explanations, with heatmaps, bar charts, and trend lines being the most frequently used. A wide range of other formats,such as scatter plots, histograms, Sankey diagrams, and tree structures, are also represented, reflecting the diversity of visual explanation strategies.}
\end{figure}

The presentation of explanations to users is crucial in determining the effectiveness and usability of \acp{XUI}. \Cref{tab:explanation_modalities} categorises the explanation modalities observed in the surveyed studies. Visual explanations dominate, likely due to their intuitive and accessible nature. Formats such as graphs, heatmaps, and diagrams translate complex AI reasoning into interpretable insights for users. Textual explanations are also widely adopted, reflecting the value of concise descriptions for clarity. Natural language explanations, audio, and tangible formats are comparatively underutilised, potentially reflecting their more specialised use cases or higher development costs. Video explanations are rare, possibly due to the time-intensive nature of their production and the immediate interactivity required in many \acp{XUI}. \Cref{fig:visual_tax_bar} details the visualization types employed in \acp{XUI}. Heatmaps stand out as the most frequently used visualization, likely due to their ability to quickly convey patterns or feature attributions \revi{and their widespread use in image explanations in general}. Bar charts and trend lines are common, reflecting their utility in comparative and temporal data analysis. Hybrid approaches that integrate textual and visual elements seem to be preferred in the literature, suggesting that combining modalities may provide the most effective explanations by complementing graphical clarity with textual context.

\subsubsection{\revi{Interactivity}}\label{sec:rq2-interactivity}
Interactivity is a fundamental aspect of \acp{XUI}, influencing how users engage with and interpret \ac{AI} explanations. Across the surveyed studies, we identified 95 interactive \acp{UI} and 47 non-interactive ones.

Interactive \acp{XUI} allow users to manipulate explanation parameters, explore different perspectives, or request additional details, potentially enhancing comprehension and trust \cite{Weitz2021Let,Gromowski2020Process}. These interfaces often employ dynamic visualization techniques, adjustable feature importance rankings, or interactive counterfactual explanations, enabling users to tailor explanations to their needs. Conversely, non-interactive \acp{XUI} present static explanations without user input \cite{Riveiro2021Thats,Upasane2021Big}. While these may still provide valuable insights, they often lack adaptability, potentially limiting their effectiveness in complex decision-making scenarios. The prevalence of interactive systems in the dataset suggests a growing emphasis on user engagement in XAI research, highlighting the importance of designing explanations that support exploratory and user-driven interactions.

\subsection{\texorpdfstring{RQ3. What solutions are used to evaluate \acp{XUI}?}{RQ3. What solutions are used to evaluate XUIs?}}\label{sec:RQ3}

This research question \revi{is characterized from a human-centered perspective, since it} explores the relationship between the user, the type of user study conducted, and the \revi{evaluation constructs (e.g., usability, trust)} employed, bridging the Human-Computer Interaction perspective with the algorithmic approach to offer a comprehensive understanding of the \acp{XUI} environment. In the development of \acp{XUI}, evaluating how users interact with and perceive these systems is critical to ensuring their effectiveness, usability, and trustworthiness \cite{DBLP:conf/chi/AbdulVWLK18}. The way users are evaluated and the \revi{constructs} observed play a central role in understanding whether the explanations provided by an \ac{AI} system align with user needs and expectations. By examining study methodologies and key performance metrics, researchers can identify strengths and gaps in \ac{XUI} design, ensuring that these systems are technically robust and user-centered. This analysis is essential for advancing the field of \ac{XAI} and fostering greater adoption of \ac{AI} systems in real-world applications where transparency and trust are \revi{crucial}.\revi{\paragraph{Summary of findings} The most common evaluation methods used with \acp{XUI} are controlled experiments (typically used with domain experts and non-experts), followed by interviews (often used with domain experts, capturing qualitative data) and usablity studies. Domain experts are typically involved in evaluating \acp{XUI} in high-stakes domain (e.g., health and finance), while \ac{AI} experts are also involved in usability studies and interactive feedback and co-design sessions. Evaluation studies often measure trust, usability, workload, perceived effectiveness, and satisfaction. Constructs such as helpfulness are typically measured with domain experts, while \ac{AI} experts are generally involved in studies measuring task performance, usability, and confidence.}

\begin{table*}[t]
    \centering
    \small 
    
    \begin{minipage}[t]{0.25\textwidth}
        \centering
        \caption{Expl. Modalities}
        \label{tab:explanation_modalities}
        \begin{tabular}{@{} lc @{}}
            \toprule
            Modality & Count \\ \midrule
            Visual & 76 \\
            Text & 50 \\
            Natural lang. & 6 \\
            Audio & 3 \\
            Tangible & 2 \\
            Video & 1 \\
            Other & 2 \\
            \bottomrule
        \end{tabular}
    \end{minipage}%
    \hfill 
    \begin{minipage}[t]{0.36\textwidth}
        \centering
        \caption{Type of User study}
        \label{tab:user_study}
        \begin{tabular}{@{} lc @{}}
            \toprule
            User Study & Count \\ \midrule
            Controlled Exp. & 53 \\
            Interview & 28 \\
            User Observation & 18 \\
            Usability Study & 15 \\
            Survey & 10 \\
            Focus-Group & 4 \\
            Interactive session & 4 \\ 
            Wizard-of-Oz & 2 \\
            UI inspection & 2 \\
            \bottomrule
        \end{tabular}
    \end{minipage}%
    \hfill 
    \begin{minipage}[t]{0.35\textwidth}
        \centering
        \caption{Top 14 Eval. \revi{Constructs}}
        \label{tab:user_metrics}
        \begin{tabular}{@{} lc @{}}
            \toprule
            Eval. \revi{Constructs} & Count \\ \midrule
            Trust & 33 \\
            Understandability & 30 \\
            Usability & 24 \\
            Satisfaction & 20 \\
            Usefulness & 15 \\
            Workload & 15 \\
            Perceived Effec. & 14 \\ 
            Task Perform. & 14 \\
            Transparency & 8 \\
            Perc. Quality & 8 \\
            Helpfulness & 8 \\
            Confidence & 7 \\
            Ease of use & 6 \\
            Explainability & 6 \\
            \bottomrule
        \end{tabular}
    \end{minipage}
\end{table*}
\subsubsection{\revi{Evaluation Methods and Constructs in XUI Research}
\Cref{tab:user_study}} categorises the various study methods used to evaluate XUIs. Controlled experiments are the most frequent ones, highlighting the importance of rigorous, systematic evaluation under controlled conditions. This method allows researchers to isolate specific factors and measure their impact on user interaction with XUIs. Additionally, interviews and usability studies are frequently utilised, highlighting the focus on user perspectives and ensuring that the interfaces remain practical and accessible.

Moving onto the \revi{evaluation constructs,} illustrated in \cref{tab:user_metrics}, \textit{trust} stands out as central, indicating that fostering user confidence in AI systems is a key priority for XUI developers. \textit{Usability} and \textit{workload} are close behind, emphasizing the importance of ensuring the interface is easy to use and does not overwhelm the user. \textit{Perceived effectiveness} and \textit{satisfaction} suggest an interest in understanding users' perceptions of how well the system supports their needs. \Cref{tab:metrics_refs} shows the most popular \revi{evaluation constructs}. Collectively, these constructs underscore the importance of designing XUIs that not only enhance explainability but also encourage practical adoption in real-world applications. \acp{XUI} should, in fact, be designed to inspire trust, remain user-friendly, and avoid overwhelming users with excessive complexity.

It is worth noting that, despite the importance of transparency in explainability research, \revi{it} has only been explicitly assessed in eight studies, suggesting that current research in \acp{XUI} should place greater emphasis on incorporating transparency \revi{constructs} into evaluations.

\begin{table}[t]
    \centering
    \small
    \caption{Papers that report user studies evaluating Trust, Usability, and Workload}
    \label{tab:metrics_refs}
    \resizebox{\linewidth}{!}{%
    \begin{tabular}{@{} l c l @{}}
        \toprule
        Evaluation Constructs & N. Papers & References \\ 
        \midrule
        Trust & 28 & \cite{Weitz2021Let, Kim2023AlphaDAPR, Wysocki2023Assessing, Guo2022Building, Khurana2021ChatrEx, Lee2020CoDesign, Panigutti2023Codesign, DBLP:journals/tlt/LuWCZ24, Karran2022Designing, Bhattacharya2023Directive, MezaMartinez2023Does, Hernandez-Bocanegra2021Effects, DBLP:conf/chi/BhattacharyaEXMOS2024, DBLP:conf/hci/Kim2023wheater, DBLP:conf/aime/RohrlMLKHKSHD23, Guesmi2022Explaining, DBLP:conf/hci/JoshiGKB24, Lombardi2024Exploring, Naiseh2023How, bahel2024Initial, Guesmi2024Interactive, Lee2023LIMEADE, perlmutter2024impact, DBLP:conf/fat/BertrandEM23, Park2022Reinforcement, Zytek2022Sibyl, Purificato2023Use, Heimerl2022Unraveling, NimmoUserCharacteristics2024, DBLP:conf/hci/LeiHZ24, desolda2023explanations, stites2021sage} \\
        \addlinespace 
        Usability & 24 & \cite{Fujiwara2020Visual, Wu2022AIDriven, DBLP:conf/caise/FusslNH24, Khurana2021ChatrEx, Wang2021CNN,Huang2022ConceptExplainer,Cheng2021DECE,DBLP:journals/ijmms/KimCPPNJL23, szymanski2024designing, Hohman2019Gamut, Guesmi2024Interactive, Wang2022M2Lens, Ming2019RuleMatrix, Sevastjanova2021QuestionComb, Park2022Reinforcement, DBLP:conf/chi/Zhang24Rethinking, El-Zanfaly2023Sandintheloop, Sciascio2017Supporting, Purificato2023Use, Hao23Time, Yuan24TRIVEA, Antar24VIME,Szymanski2021Visual} \\
        \addlinespace
        Workload & 15 & \cite{Fujiwara2020Visual, Piorkowski23AIMEE, Dominguez2020Algorithmic, Sanneman2022Empirical, Lee2020CoDesign, Wang2023DeepSeer, Karran2022Designing, Khanna2022Finding, Anderson2020Mental, El-Zanfaly2023Sandintheloop, Kridalukmana2022SelfExplaining, Sciascio2017Supporting, Nakao2022Involving, Heimerl2022Unraveling, Yuan2024Visual} \\
        \bottomrule
    \end{tabular}
    }
\end{table}

\begin{figure}[t]
  \centering
  \includegraphics[width=0.7\textwidth]{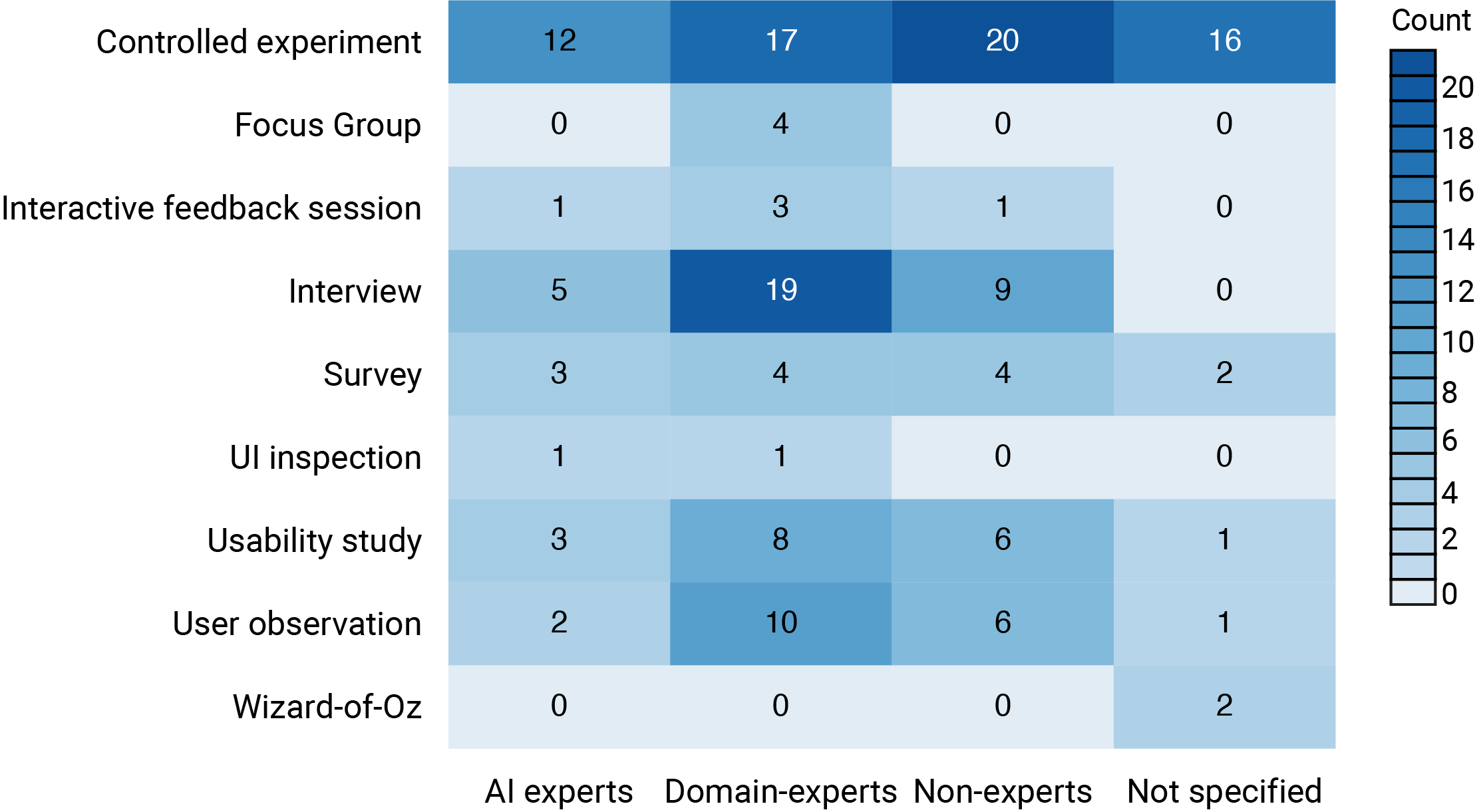}
  \caption{Methods used to evaluate explanation interfaces across different user types. The heatmap shows the distribution of evaluation methods, categorised by the type of user involved (e.g., AI experts, domain experts, non-experts).}
  \label{fig:heatmap_user_study}
  \Description{Methods used to evaluate explanation interfaces across different user types. The heatmap shows the distribution of evaluation methods such as controlled experiments, interviews, surveys, and usability studies, categorised by the type of user involved (e.g., AI experts, domain experts, non-experts). Controlled experiments and interviews are the most common methods, particularly among non-experts and domain experts, reflecting a strong emphasis on both empirical validation and qualitative feedback in XUIs evaluation.}
\end{figure}

\subsubsection{\revi{Tailoring XUI Evaluation Methods to User Groups}}
\Cref{fig:heatmap_user_study} represents the distribution of user types across various study methodologies used in the evaluation of \acp{XUI}. The figure provides some insights into how different user groups are targeted and about the methodological approaches used to assess their interaction with and understanding of the interface. The choice of study method is influenced by the target user group: controlled experiments are favoured for domain experts and lay users due to their rigorous structure. Interviews, however, are particularly prominent in studies involving domain experts, as they can usually capture qualitative insights. Domain experts are typically involved in interviews for the application of XUI technologies in the high stakes domains such as Health \cite{Hwang2022Clinical, Kim2023AlphaDAPR, kim2023designing, Bhattacharya2023Directive,DBLP:conf/chi/BhattacharyaEXMOS2024,Naiseh2023How,DBLP:journals/pacmhci/Okoloeasy24,DBLP:journals/tiis/LarasatiLM23,Lindvall2021Rapid,DBLP:conf/chi/Zhang24Rethinking,Cheng2022VBridge} and Finance and economics,  \cite{DBLP:conf/caise/FusslNH24, Purificato2023Use} but also in other domain such as Education \cite{DBLP:journals/tlt/LuWCZ24}, Natural Language Processing \cite{Wang2022M2Lens} and weather forecasting \cite{DBLP:conf/hci/Kim2023wheater}. Methods such as interactive feedback sessions, co-design, usability studies, and, notably, surveys are also employed to evaluate AI experts in various domains, such as health \cite{Hohman2019Gamut} associated with Artificial Intelligence \cite{Baniecki2024Grammar}, education \cite{Shin2022XDesign, Yuan24TRIVEA} and finance and economics \cite{Hohman2019Gamut, MezaMartinez2023Does}. While these approaches are often associated with assessing non-expert users, their application to AI experts suggests an interest in gathering qualitative and usability-focused insights from technically proficient individuals.

Notably, certain study types, such as Wizard-of-Oz \cite{Hernandez-Bocanegra2023Explaining, DBLP:conf/hci/JoshiGKB24} experiments, exhibit a high proportion of unspecified user roles, indicating that these evaluations often prioritise AI behaviour simulation over explicitly defined user categories. 

\subsubsection{\revi{User group and Evaluation Constructs}}

\begin{figure}[t]
    \centering
    \includegraphics[width=1\linewidth]{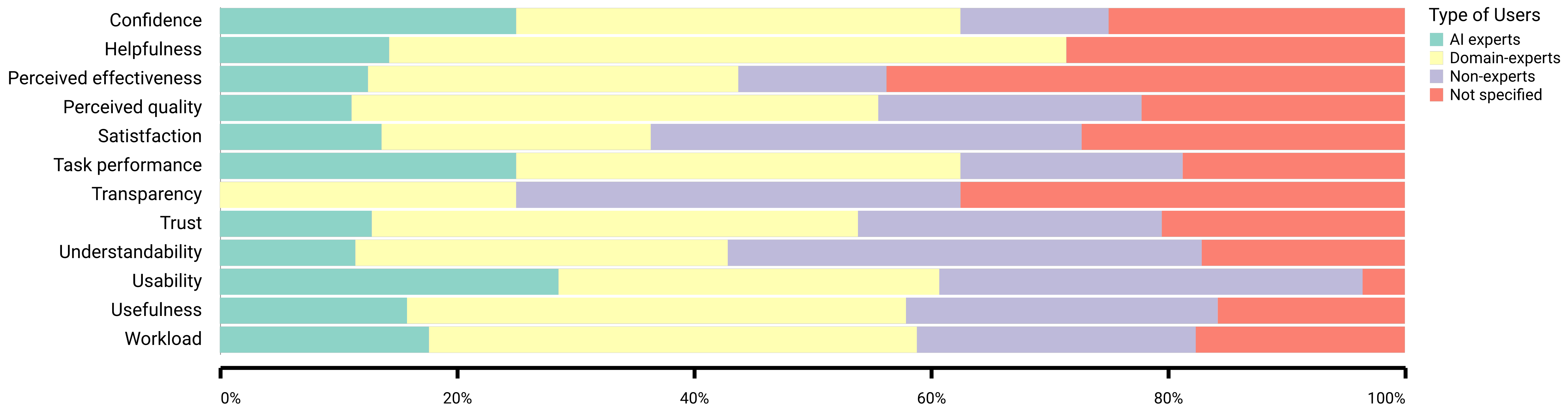}
    \caption{Reported evaluation criteria by user type. Figure shows which user groups were associated with various evaluation criteria used in assessing explanation interfaces. Criteria such as trust, understandability, usability, and perceived effectiveness were evaluated across all user types, highlighting the multidimensional nature of user-centered evaluation in explainable systems. Notably, transparency was not evaluated in studies involving AI experts.}
    \label{fig:stackedbar_metrics_user}
    \Description{Reported evaluation criteria by user type. The stacked bar chart illustrates which user groups (AI experts, domain experts, non-experts, and unspecified) were associated with various evaluation criteria used in assessing explanation interfaces. Criteria such as trust, understandability, usability, and perceived effectiveness were evaluated across all user types, highlighting the multidimensional nature of user-centered evaluation in explainable systems. Notably, transparency was not evaluated in studies involving AI experts.}
\end{figure}

The relationship between the user type involved in the study and the \revi{constructs} assessed provides insight into the key aspects that \ac{XUI} designers prioritise when evaluating an application's suitability for real-world adoption.
\Cref{fig:stackedbar_metrics_user} reveals several key trends.
Domain experts are the predominant group assessing \textit{helpfulness} \cite{Hwang2022Clinical,Nourani2022DETOXER,Liu2022Generating,Zytek2022Sibyl}; this might indicate that XUIs are designed with expert decision-making support in mind, where explainability is expected to enhance task performance rather than merely provide technical insights.
Since domain experts are the end-users of AI systems in high-stakes applications (e.g., healthcare, finance), their evaluation of helpfulness is crucial in determining whether explanations effectively aid decision-making rather than just offering transparency. Interestingly, AI experts are almost absent in studies assessing \textit{transparency}, despite their technical expertise. This suggests that transparency is being evaluated primarily from a user-facing perspective rather than from an algorithmic interpretability standpoint.
This \revi{construct} frequently appears in the Conversational AI domain \cite{Khurana2021ChatrEx,Hernandez-Bocanegra2023Explaining} and recommendation systems \cite{Hernandez-Bocanegra2021Effects,Guesmi2022Explaining, Lee2023LIMEADE, Park2022Reinforcement}, highlighting its importance in ensuring domain experts and lay-users understand system decisions and interactions in these applications. AI experts are more involved in studies that assess \textit{task performance}, \textit{usability}, and \textit{confidence}: task performance is typically assessed in applications within computer science-specific domains, such as networking \cite{Fujiwara2020Visual,Xuan2022VACCNN} and machine learning \cite{Piorkowski23AIMEE}. Usability studies involving AI experts span a broad range of domains, including mobility \cite{Antar24VIME}, education \cite{Yuan24TRIVEA}, economics \cite{Purificato2023Use,Hohman2019Gamut}, and agnostic applications \cite{Cheng2021DECE,Hao23Time}, reflecting a general need to ensure the usefulness of XUI applications. The \textit{confidence} \revi{constructs} \cite{Karran2022Designing,Baniecki2024Grammar,Zhou2021Videobased}, measured significantly less frequently compared to the others, indicates a concern for how well the system supports trust in its outputs, ensuring that explanations are not only technically accurate but also foster reliability and acceptance in real-world applications.

\subsection{\texorpdfstring{RQ4. What can guide the design of \acp{XUI}?}{RQ4. What can guide the design of XUIs?}} \label{sec:RQ4}



Several papers in our review propose guidelines for designing XUIs. These emerged either as lessons learned from evaluating XUIs within a user study or as indications of a broader framework.
From the manual analysis of the retrieved papers, \textbf{80} of them present a XUI and indicate clear design guidelines, while \textbf{10} solely present a theoretical or conceptual framework that can guide the design of XUIs. 

\Cref{tab:design_guidelines_refs} reports the list of papers that present design guidelines. We grouped the papers that propose an XUI according to the AI model used by the system. This categorization is meant to provide readers with a quick reference for the design of XUIs for systems that use a specific technology (i.e., \textit{Neural networks}, \textit{Ensemble methods}, \textit{Reinforcement Learning}, \textit{Transparent Models}, \textit{Generic AI models}, \textit{Other models}). Also, the list of papers that only present a framework for designing XUIs is reported.

\revi{\paragraph{Summary of findings} \acp{XUI} should be designed using a user-centered approach, accommodating users' mental models and favouring transparency and trust. Providing interactivity and control of explanations, providing multiple levels of visualizations, together with contextual information to better understand them, can help in this regard, as well as allowing users to personalize explanations based on their profile. Finally, popular frameworks can guide the design of \acp{XUI} by defining how explanations should be shaped (e.g., which questions they should answer), what values a \ac{XUI} system should possess (e.g., transparency, engagement, control on interactivity), how users may be involved in the design process, and how to align user goals with system explanations. 
}

\begin{table}[t]
    \centering
    \small
    \caption{\revi{Papers that report design guidelines divided by AI model}}
    \label{tab:design_guidelines_refs}
    \begin{tabular}{@{} p{.23\linewidth} p{.13\linewidth} p{.59\linewidth} @{}}
        \toprule
        \revi{AI model} & \revi{Num. papers} & \revi{References} \\ \midrule
        \revi{Neural networks} & \revi{29} & \cite{Weitz2021Let, Hwang2022Clinical, Fujiwara2020Visual, Krause2017Workflow, Dominguez2020Algorithmic, Kim2023AlphaDAPR, Wang2021CNN, Lee2020CoDesign, Huang2022ConceptExplainer, Malandri2023ConvXAI, Wang2023DeepSeer, kim2023designing, Nourani2022DETOXER, MezaMartinez2023Does, Hernandez-Bocanegra2021Effects, DBLP:conf/hci/Kim2023wheater, Guesmi2024Interactive, Lee2023LIMEADE, Wang2022M2Lens, Esfahani2024Preference, Lindvall2021Rapid, DBLP:conf/chi/Zhang24Rethinking, Hao23Time, Heimerl2022Unraveling, DBLP:journals/vi/ChotisarnGZC23, Gehrmann2019Visual, Mishra2022Why, Ming2019RuleMatrix, Karran2022Designing} \\  
        \revi{Ensemble methods} & \revi{9} &
        \cite{Wysocki2023Assessing, Chromik2021Think, Wang2019Designing, Cheng2022VBridge, DBLP:conf/chi/BhattacharyaEXMOS2024, Bove2023Investigating, Ma2022Explainable, Kalamaras2019Visual, Cau2023Supporting} \\
        \revi{Reinforcement Learning} & \revi{3} &
        \cite{Khanna2022Finding, Park2022Reinforcement, Mishra2022Why} \\
        \revi{Transparent models} & \revi{5} &
        \cite{Guo2022Building, Nakao2022Involving, Bhattacharya2023Directive, Hohman2019Gamut, Lee2020CoDesign} \\
        \revi{Generic/Model-agnostic models} & \revi{30} &
        \cite{Ferrario2020ALEEDSA, Sanneman2022Empirical, Cabour2023Explanation, Su2020Analyzing, Naiseh2024cxai, mosca2021elvira, fiok2022education, Raees2024Explainable, Liu2022Generating, Laato2022How, DBLP:conf/nordichi/MeyerZ24, Sanneman2022Situation, Ferreira2020What, Simkute2022XAI, Wu2022AIDriven, DBLP:conf/caise/FusslNH24, Khurana2021ChatrEx, DBLP:conf/chi/JansenLWZ24, Cheng2021DECE, szymanski2024designing, Schneider2021ExplAIn, Collaris2020ExplainExplore, Kim2023Help, Naiseh2023How, DBLP:journals/pacmhci/Okoloeasy24, bahel2024Initial, DBLP:journals/tiis/LarasatiLM23, Zytek2022Sibyl, Szymanski2021Visual, DBLP:conf/hci/LeiHZ24}
        \\
        \revi{Other AI models} & \revi{4} &
        \cite{Riveiro2021Thats, Bove2022Contextualization, Piorkowski23AIMEE, desolda2023explanations}
        \\ 
        \bottomrule
        \revi{Frameworks only} & \revi{10} &
        \cite{Cabour2023Explanation, Su2020Analyzing, Naiseh2024cxai, fiok2022education, DBLP:conf/nordichi/MeyerZ24, Sanneman2022Situation, Simkute2022XAI, Raees2024Explainable, Liao2020Questioning, pieters2011explanation}
        \\
        \bottomrule
    \end{tabular}
\end{table}

\subsubsection{\revi{Core design principles for XUIs}}
Our review of XUI design guidelines across various AI models---including Neural networks, Ensemble methods, Reinforcement learning, Transparent models, Generic approaches, and other AI models---reveals several cross-cutting principles. These core principles serve as a foundation for effective XUI design, even as model-specific requirements require tailored adaptations. In the following, we discuss these principles, also illustrating them within different AI approaches. 

\paragraph{User-Centered Design}
At the heart of any effective XUI is a design that is grounded in the user's needs, context, and expertise. 
At a high level, explanations in XUIs should use clear, jargon-free language that adapts to the user's context, reducing cognitive load and making AI behaviour more accessible \cite{Khurana2021ChatrEx, DBLP:conf/chi/JansenLWZ24, DBLP:journals/pacmhci/Okoloeasy24, szymanski2024designing}. 
User-centered frameworks such as \cite{Barda2020Qualitative} should be leveraged to define the purpose, content, and presentation of explanations in contexts such as clinical decision support \cite{Hwang2022Clinical}. Ensemble-based XUIs emphasise the need to account for users' mental models and to minimise cognitive load \cite{Wysocki2023Assessing, Chromik2021Think}, while generic systems advocate for tailoring explanations to both novice and expert users \cite{Khurana2021ChatrEx, DBLP:conf/chi/JansenLWZ24}. Particular effort should be put into this aspect by researchers (e.g., using participatory design, conducting usability studies), as often solutions that appear effective to developers are not ideal to end-users \cite{Guo2022Building}.

\paragraph{Interactivity and Control}
Effective \acp{XUI} empower users by supporting active engagement with the system. Supporting interactive exploration was found useful for users, for example, allowing selection, filtering, juxtaposition, and smooth transitions between views \cite{Wang2021CNN, Hao23Time}. 
Moreover, features such as what---if exploration, guided navigation, and real-time updates were found to enhance user engagement and comprehension \cite{Collaris2020ExplainExplore, Liu2022Generating}. 
Interaction is also a core requirement for white-box models to iteratively refine the understanding of explanations \cite{Guo2022Building, Nakao2022Involving}. 
Regarding reinforcement learning systems, interaction modalities such as questioning the model and navigating through explanation spaces help users clarify why actions were taken, increasing trust and understanding of the system \cite{Mishra2022Why}. 

\paragraph{Transparency and Trust}
User trust is fostered by a high level of transparency of the AI's decision-making process. To build trust, the potential and limitations of the AI system should be properly conveyed by ensuring that users receive accurate cues about model confidence and reliability \cite{DBLP:journals/pacmhci/Okoloeasy24, Zytek2022Sibyl}. 
With black-box models such as neural networks, transparency can be achieved by displaying detailed insights, such as model uncertainties, confidence scores, and error metrics \cite{Lee2023LIMEADE, Malandri2023ConvXAI, Nourani2022DETOXER, DBLP:conf/chi/Zhang24Rethinking}. Offering descriptive statistics and summaries of model performance can allow users to gauge overall system reliability \cite{Lee2023LIMEADE, Malandri2023ConvXAI}. White-box systems, in particular, stress that explanations should faithfully represent how the underlying model functions, including both local and global insights \cite{Nakao2022Involving, Guo2022Building}, possibly to increase understanding of the risk behind a model's decision \cite{Bhattacharya2023Directive}. To increase trust, explanations should use an adequate language that is either accessible and avoids technical jargon \cite{DBLP:journals/pacmhci/Okoloeasy24} or is tailored to the user's domain and cultural context \cite{szymanski2024designing}. Explanations should also resemble human reasoning and the way of explaining \cite{Kim2023Help}. Finally, special care should also be taken in the selection of features to be addressed in explanations, as not every feature may be understandable and meaningful to the end user \cite{desolda2023explanations}. 

\paragraph{Multi-level Visualizations}
Effective visual representations in XUIs are essential for helping users understand complex AI behavior. Offering visual summaries through multi-level adjacent explanation visualizations was found to improve user confidence \cite{Karran2022Designing}. Transitions and animations between different visualizations can help navigate the XUI, improve understanding of the underlying AI model, and increase learning engagement and enjoyment \cite{Wang2021CNN}. In reinforcement learning, an effective visualization of state spaces should provide both high-level overviews and detailed insights into specific states \cite{Mishra2022Why}. XUIs can also benefit from visualizations such as color-coded risk factors, feature importance charts, and graphical comparisons that clarify both instance-specific and overall model behavior, also by showing counterfactuals and similar instances \cite{Bhattacharya2023Directive, Hohman2019Gamut, Lee2020CoDesign}. 

\paragraph{Contextual Information}
Placing explanations together with additional contextual information enhances their relevance and interpretability to domain-expert users, e.g., physicians \cite{Hwang2022Clinical,  Cheng2022VBridge}, AI engineers \cite{Fujiwara2020Visual, DBLP:conf/chi/BhattacharyaEXMOS2024}, etc. Moreover, the depth of information in explanations should be adaptable to the user's expertise, ensuring that both detailed and summary information are available without overwhelming the user \cite{Wang2019Designing, Naiseh2023How, DBLP:conf/hci/LeiHZ24}.

\paragraph{Personalization and Adaptability of Explanations} 
The results from the literature stress the need for culturally sensitive and user-tailored content, ensuring that explanations are accessible and meaningful for diverse audiences \cite{DBLP:journals/pacmhci/Okoloeasy24, bahel2024Initial}. Designs should tailor content to the user's background, cognitive abilities, and cultural context. This includes strategies such as incremental disclosure of information and the use of both textual and visual elements to match different learning styles \cite{Kim2023Help, Naiseh2023How}.

\subsubsection{\revi{Frameworks}} 

\revi{The ten frameworks resulting from the literature review cover heterogeneous aspects of the design of \acp{XUI} and may be grouped into three categories:} 
\revi{
\begin{enumerate*}
    \item \textbf{Interaction and Content Design Frameworks}: These frameworks focus on the artifact (i.e., the explanation itself) and provide operational guidance on mapping user inquiries to specific interface components and explanation techniques.
    \item \textbf{Axiological and Relational Design Frameworks}: These frameworks focus on human values and the relationship between the user and the AI, guiding the system's high-level ethical positioning and prioritizing trust/confidence calibration, user engagement, and situational awareness. 
    \item \textbf{Cognitive and Learning Alignment Frameworks}: These frameworks focus on cognitive aspects of the user, drawing upon cognitive psychology and learning theories to align the \ac{XUI} with human information processing capabilities, literacy levels, and learning goals.
\end{enumerate*}}

\paragraph{\revi{Interaction and Content Design Frameworks}}
In \cite{DBLP:conf/nordichi/MeyerZ24}, Meyer and Zhu examined 28 XUIs and linked specific UI components to explanation types. While ``Why'' and ``What-if'' questions are commonly addressed, tools for ``How'' or ``Input'' explanations are scarce, with elements like sliders, buttons, and icons playing key explanatory roles. 
\citet{Liao2020Questioning} identified gaps between XAI research and design practice by interviewing 20 practitioners, resulting in an XAI question bank based on prototypical user inquiries. These questions are shaped by user context and mapped to suitable explanation techniques, such as global or counterfactual methods.
\citet{Su2020Analyzing} analyzed 40 mobile health apps and found that users want AI features---like recommendations or predictions---explained in clear, engaging ways. The study emphasises aligning explanations with user literacy and providing meaningful feedback loops to refine AI behavior.

\paragraph{\revi{Axiological and Relational Design Frameworks}}
The Value Sensitive Design methodology \cite{Naiseh2024cxai} emphasises embedding human values---such as transparency, autonomy, and inclusivity---into explainable AI systems. Similarly, \citet{Raees2024Explainable} advocate for user engagement and transparency to enhance trust in interactive AI.
\citet{Sanneman2022Situation} proposed SAFE-AI, a framework for enhancing user situation awareness by addressing what AI did, why it did so, and what it will do. It also includes trust calibration and workload management to optimise user-AI interaction.
\revi{\citet{pieters2011explanation} distinguishes between \textit{explanation-for-confidence} and \textit{explanation-for-trust}, showing how explanation goals shape user understanding; the author argues that explanation depth must match its intent: too little detail may hinder trust, while too much may lower confidence.}

\paragraph{\revi{Cognitive and Pedagogical Alignment Frameworks}}
In education, \citet{fiok2022education} advocate for participatory, user-centered design in XAI tools. Their work highlights the importance of grounding design in psychological theory and leveraging HCI expertise to align technical systems with learners' needs and trust-building goals. 
\citet{Simkute2022XAI} presented design principles for educational XUIs based on cognitive psychology; effective learning support includes techniques such as cognitive forcing, analogies, contrastive reasoning, and interactive feedback to promote skill development.
The Abstracted Explanation Space \cite{Cabour2023Explanation} offers a structured framework that links user goals with appropriate explanation types, covering perception, comprehension, and projection. \revi{This framework} advocates cross-disciplinary collaboration and helps designers tailor explanations to avoid cognitive overload and enhance relevance.

\section{Discussion}\label{sec:hermes}

\begin{figure}[b]
    \centering
    \includegraphics[width=0.85\linewidth]{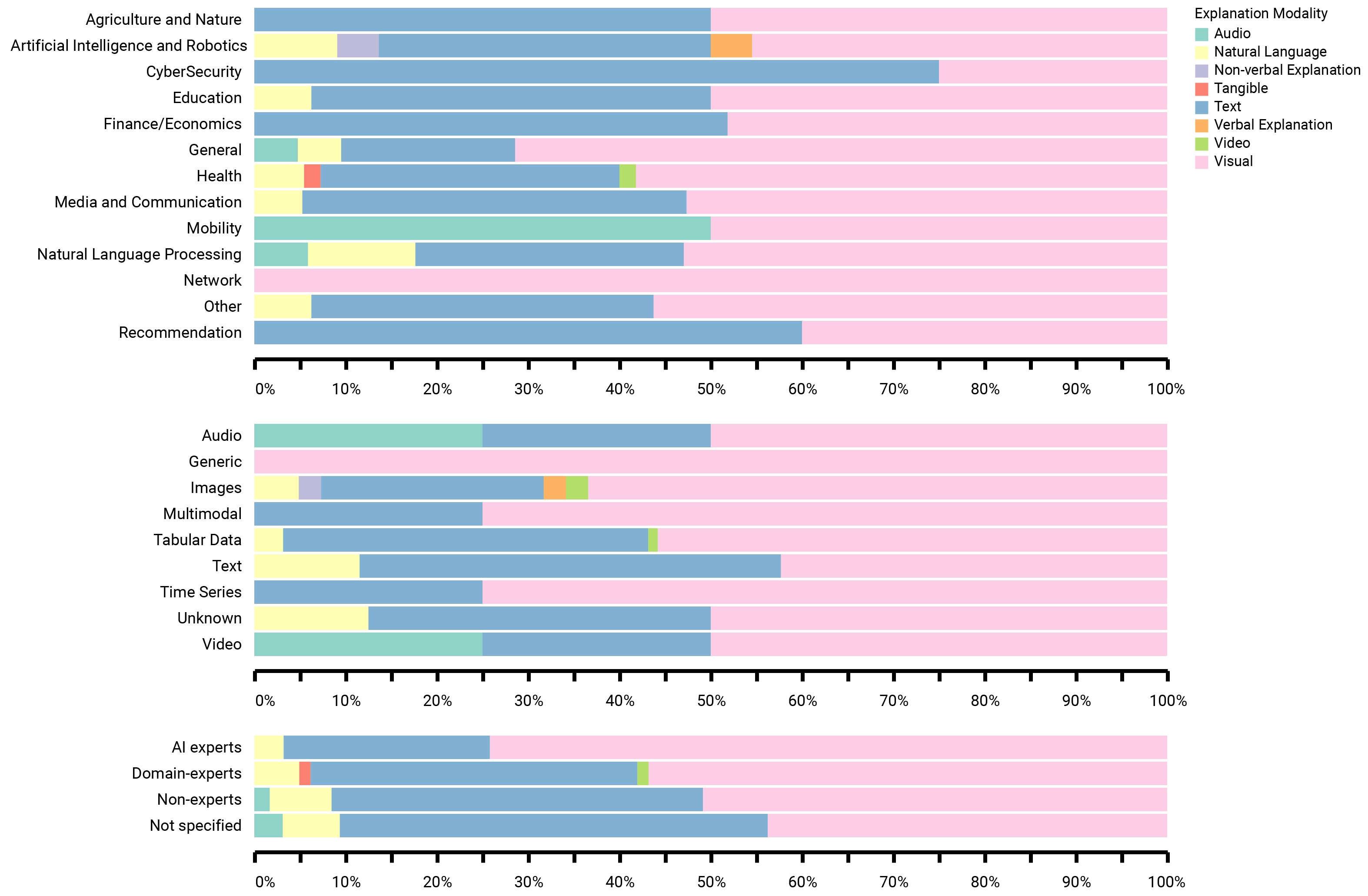}
        \caption{Distribution of explanation modalities across different application domains, data types, and types of users. Visual explanations dominate across most contexts, while other modalities, such as natural language, non-verbal, or tangible explanations, vary depending on the domain and target user group. The visualization reveals distinct preferences and trends in how explanations are tailored based on the nature of the data and the expertise of the intended users.}
        \label{fig:stacked_expl-modality_datatype}
        \Description{Distribution of explanation modalities across different application domains, data types, and types of users. The chart highlights how visual explanations dominate across most contexts, while other modalities---such as natural language, non-verbal, or tangible explanations---vary depending on the domain and target user group. This visualization reveals distinct preferences and trends in how explanations are tailored based on the nature of the data and the expertise of the intended users.}
\end{figure}

This section \revi{discusses the results} by relating the dimensions analyzed in our RQs. \revi{dditionally, to make our discussion actionable, we present a platform to support practitioners in building effective \acp{XUI}: \acs*{HERMES}.} We examine how each dimension interacts across all RQs to identify recurring patterns. This approach allows us to distill insights that can support practitioners in navigating the broader landscape of the findings.

Visual \revi{and textual explanations} are the primary means of conveying explanations, as shown in \Cref{fig:stacked_expl-modality_datatype}, where the explanation modality is shown relative to its use within different domains, types of data, and types of users. 
\revi{This} preference is particularly evident when working with tabular data and images, where graphical representations can make patterns and relationships more apparent. However, visual explanations are not limited to these data types; they are also applied to text and time series, while \revi{textual} explanations remain common for tabular data. It is also interesting to note how in the mobility domain \cite{DBLP:conf/chi/SchneiderHRGTG21}, audio is present.

Certain explanation methods are widely used across domains. In healthcare, \ac{SHAP} \cite{Barda2020Qualitative, Wysocki2023Assessing, DBLP:journals/ijmms/KimCPPNJL23, Cheng2022VBridge, Zhou2021Videobased}, feature importance \cite{Lee2020CoDesign, DBLP:conf/aime/RohrlMLKHKSHD23, DBLP:conf/hci/LeiHZ24, Mishra2022Why}, and counterfactual explanations \cite{Lamy2019Explainable, DBLP:conf/chi/Zhang24Rethinking} are frequently employed to interpret model decisions. These techniques are also common in finance and economics, where counterfactual reasoning \cite{Ferrario2020ALEEDSA, Ma2022Explainable, Sovrano2021Philosophy, Hohman2019Gamut, Bove2021Building, Esfahani2024Preference, Cau2023Supporting} and \ac{SHAP} values \cite{MezaMartinez2023Does, Chromik2021Making, Purificato2023Use} help practitioners assess risk and model fairness. Since these methods are well-suited for tabular data, they naturally find application in domains where structured datasets are prevalent.

The choice of explanation technique often depends on the audience. \ac{SHAP} \cite{Malandri2023ConvXAI, Ma2022Explainable, Chromik2021Making, He2024VMS, DBLP:conf/chi/JansenLWZ24} explanations tend to be preferred when working with non-expert users due to their intuitive visualization of feature contributions. However, exemplars i.e., concrete instances that illustrate a model's behavior, and salient masks, which highlight the most influential input regions, are also commonly used to aid understanding \cite{Huang2022ConceptExplainer, Wang2021Explainability, Yang2020How, DBLP:journals/tiis/LarasatiLM23, Nakao2022Involving}. When designing user studies for \acp{XUI}, the choice of method should align with the target user group. Interviews are most commonly used for domain experts, particularly in high-stakes fields like healthcare and finance, where qualitative insights are crucial for understanding decision-making needs. In contrast, surveys and usability studies are more prevalent among \ac{AI} experts, emphasizing practicality and efficiency in technical contexts. Regarding evaluation \revi{constructs}, domain experts in high-stakes applications tend to prioritise helpfulness \cite{clinical2022Hwang, Nourani2022DETOXER, Liu2022Generating, Zytek2022Sibyl} over transparency \cite{Sevastjanova2021QuestionComb}. Their primary concern is whether an explanation supports their decision-making process rather than simply revealing model internals. For \acp{XUI} designers, this means that explanations should be crafted to disclose system behaviour and integrate seamlessly into expert workflows, ensuring they provide meaningful, actionable insights. \revi{The \ac{SLR} presented in this paper has drawn on works from multiple disciplines to gather knowledge from diverse expertise fields. These vary substantially according to several factors, such as the application domain, the technologies underlying an \ac{AI} system, and the type of users. Therefore, accessing these findings and integrating them into a specific design context would be cumbersome and require extensive knowledge of different fields (e.g., a designer should be knowledgeable about \ac{AI} technologies). To make the findings more accessible, we developed a web platform called \acs{HERMES}, which is described in the following.}

\subsection{\revi{Navigating the Results:} \texorpdfstring{Introducing the \acs{HERMES} Platform}{Introducing the HERMES Platform}}
 
\begin{figure}[t]
    \centering
    \includegraphics[width=0.45\linewidth, alt={\ac{HERMES}}]{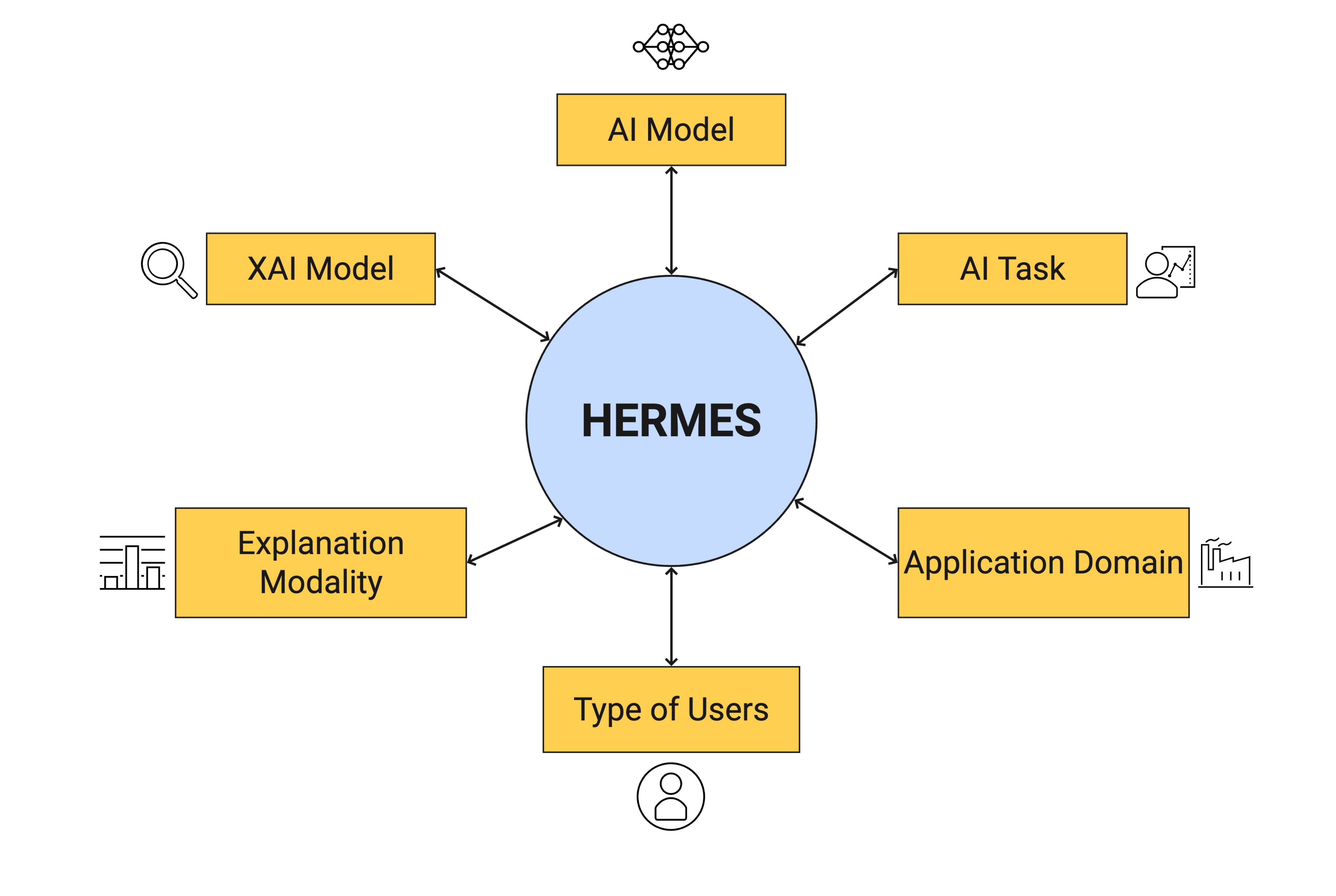}
    \caption{The HERMES Framework with the 6 dimensions observed in our analysis.}
    \label{fig:HERMES}
    \Description{The HERMES Framework with the six dimensions observed in our analysis}
\end{figure}

\begin{figure}[b]
    \centering
    \includegraphics[width=0.6\linewidth]{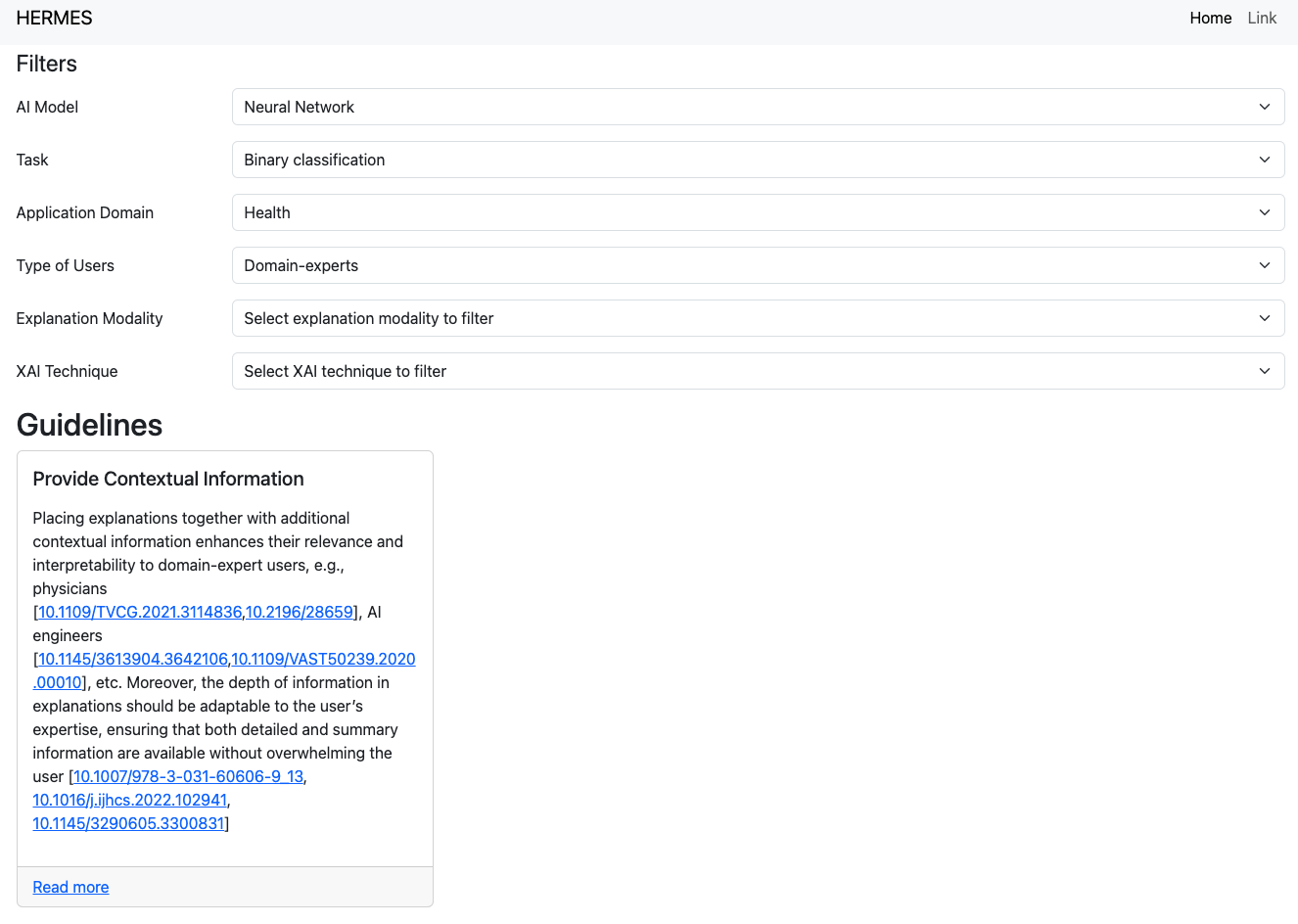}
    \caption{A screenshot of the HERMES platform showing guidelines for specific defined criteria: (\textit{AI Model} = ``Neural network'', \textit{AI Task} = ``Binary classification'', \textit{Application Domain} = ``Health'', and \textit{Type of Users} = ``Domain-experts''.}
    \label{fig:HERMES_example}
    \Description{A screenshot of the HERMES platform showing guidelines for specific defined criteria: ({AI Model} = ``Neural network'', {AI Task} = ``Binary classification'', {Application Domain} = ``Health'', and {Type of Users} = ``Domain-experts''.}
\end{figure}

\acs{HERMES} (\acl{HERMES}) is a web platform designed to support practitioners in building effective \acp{XUI}. \revi{It} offers a structured collection of design guidelines derived from the \revi{results of the \ac{SLR}}, presented as interactive cards that synthesize insights from the referenced papers. Each card includes a concise description of a guideline, relevant tags that reflect key dimensions from the associated paper, and a direct link to the original source. Users can explore the guidelines in two complementary ways: by browsing the full collection of cards, or by filtering \revi{them} through up to six entry points (\cref{fig:HERMES}), which correspond to six of the thirteen dimensions of this survey (see \cref{tab:research_dimensions}): (1) Application Domain, (2) User type, (3) AI model, (4) AI task, (5) XAI model, and (6) Explanation modality. \acs{HERMES} was developed using Jekyll\footnote{https://jekyllrb.com/} and enables designers to either align their \acp{XUI} with an existing use context \revi{(by analyzing the above dimensions)} or to explore potential design directions under specific project constraints, fostering both conformity and creativity in the design process.
Not all the input dimensions must be defined by the designer. Moreover, the output of the platform also includes data about the techniques to employ for the evaluation of the XUI (i.e., the \textit{Type of Study} to conduct and the \revi{\textit{evaluation constructs} to assess)}.  
The platform is available at the following address: \url{http://espositoandrea.github.io/hermes-mirror/}. A screenshot of the web platform is reported in \cref{fig:HERMES_example}.

\subsection{Use case: a medical application}

\revi{Let us suppose that a} medical technology company is developing a \ac{XUI} for an existing AI-based system designed to assist clinicians in diagnosing cervical cancer using MRI scans. The existing software that currently helps clinicians is a neural network-based system that determines whether a patient's blood test suggests that the clinical criteria for diabetes are met. However, the \ac{AI} model only gives the clinician a binary outcome (yes/no), without providing an explanation. The development team is thus tasked to implement an \ac{XAI} model and visualise the outputs for clinicians, who need transparent and interpretable AI outputs to support their diagnostic decisions.

To guide the design of this \ac{XUI}, the development team uses \acs{HERMES} to determine the most suitable explanation and \ac{UI} design strategies by setting four out of the six entry points as project constraints: \textit{Health} as the application domain, \textit{Neural Network} as the \ac{AI} model, \textit{Image classification} (cancerous vs. non-cancerous MRI scans) as the \ac{AI} task, and \textit{\revi{Domain} Expert} as the user type, as the interface is targeted to domain experts (i.e., clinicians). 
\revi{Since there are no other design constraints, the other two entry points (\ac{XAI} model and Explanation modality) are left empty -- thus becoming suggestions from \acs{HERMES}.}
Based on the provided parameters, \acs{HERMES} suggests explanation methods that have been effective in similar medical applications. 
\revi{The system highlights several \ac{XAI} techniques as suitable, including Counter-exemplars/factual and Decision Rules, and Natural Language, Text, and Visual explanations as explanation modalities. \acs{HERMES} also identifies the design guideline \textit{Provide Contextual Information}, which states that placing explanations alongside additional contextual information enhances their relevance and interpretability for domain-expert users, such as physicians, while tailoring the depth of explanations to the user's needs.}
To support this recommendation, \acs{HERMES} provides references to relevant academic sources, including a specific paper by \citet{Cheng2022VBridge}. In the paper, the developers can find more implementation details and access a visual example to guide the design of the new \ac{XUI}.
To ensure the \ac{XUI} meets clinicians' needs, \acs{HERMES} suggests that user studies for designing and evaluating such interfaces should include \textit{interviews} and \textit{user observation} studies. This may be interpreted as a recommendation to conduct interviews with clinicians early in the design process, and then use observational studies to iteratively test and refine the \ac{XUI}, following a human-centered design approach \cite{ISO20199241210}.
Using \acs{HERMES}, the design team efficiently selects an adequate explanation approach tailored to expert clinicians in a medical domain. The platform informs the selection of \ac{XAI} models and explanation modalities, and also suggests a human-centered approach to involve end-users both in the design and evaluation of the \ac{XUI}. This structured methodology can indeed accelerate the design process, reducing trial-and-error efforts and ensuring that the final interface aligns with best practices found in the academic literature.

\section{Future Challenges}\label{sec:challenges}
Among the future challenges identified by this \ac{SLR}, a key issue is the limited adoption of co-design practices and focus group methodologies in the development of \ac{HCXAI} systems. These approaches are essential to strengthen a real co-creation dynamic between users and \ac{XAI} systems, ensuring that user needs, mental models, and expectations are meaningfully integrated into the design process. Similarly, the collection of quantitative data and the execution of usability studies involving both \ac{AI} and domain experts are critical for increasing the practical applicability and reliability of these tools. Recent research acknowledges the growing role of \acp{LLM} in \ac{XAI}. \citeauthor{DBLP:conf/hci/LeiHZ24} suggest that with the rising adoption of LLMs, the demand for XAI among non-expert users will increase. They propose a set of design principles for \acp{XUI} \cite{DBLP:conf/hci/LeiHZ24}. \acp{LLM} are also beginning to be incorporated into Explainable Conversational Interfaces. For example, \citeauthor{DBLP:conf/hci/JoshiGKB24} conducted a Wizard of Oz user study comparing two versions of a vacation planning chatbot---one with low explainability and one with high explainability---involving 60 participants. Their study finds that providing explanations enhances trust and acceptance of the \ac{LLM}-based system \cite{DBLP:conf/hci/JoshiGKB24}, \acp{LLM} represent both a novel challenge and a promising opportunity for the development of \acp{XUI}. They are reshaping user-explanation interactions and advancing the potential for more conversational, human-like explanations, as envisioned by \citeauthor{Miller2019Explanation} \cite{Miller2019Explanation}. However, LLMs themselves remain opaque, prone to hallucinations, and can undermine user trust due to their lack of transparency and controllability. Addressing these matters is key for the future of effective and responsible \acp{XUI} design.

A significant future challenge for the  \acs{HERMES} platform lies in the development and validation of more specific and context-aware design guidelines, particularly those that intersect multiple design dimensions. Currently, many guidelines remain broad or generic, limiting their applicability to complex or highly specialised use cases. Advancing \acs{HERMES} will therefore require the formulation of more granular recommendations tailored to particular combinations of application domains, user types, AI tasks, and explanation modalities. To achieve this, integrating crowd-sourcing mechanisms could play a crucial role. By involving a diverse community of practitioners and researchers, it would be possible to both enrich the guideline repository with real-world insights and validate existing guidelines through collective evaluation. This participatory approach could significantly enhance the relevance, reliability, and adaptability of the platform, making it a more robust decision-support tool for the design of \acp{XUI} across varied contexts.


\section{Limitations and Threats To Validity}\label{sec:limitations}
In general, several threats to validity can undermine the results of a \acl{SLR}. In the following section, we report the most common ones, detailing how we mitigated them. \noindent \textbf{Selection Bias:} This occurs when the studies included in the review are not representative of the entire population of studies on the topic. This has been mitigated by: i) manually reviewing the publications to ensure their compliance with the \ac{SLR} goal, and ii) performing two phases, i.e., search on digital libraries and snowballing. However, our search focused only on full publications; therefore, we acknowledge that this \ac{SLR} does not cover demos, posters, and working papers presented at workshops. 
\textbf{Publication Bias:} This occurs when studies that show statistically significant results are more likely to be published than studies that do not. This aspect has been mitigated by manually reading those publications that do not report any results but only a technical solution with preliminary results. Besides the generic inclusion criteria, their relevance for our \ac{SLR} is considered, for example, the number of citations and the novelty of the solution. 
\textbf{Time Lag Bias:} This occurs when the review does not include all relevant studies because they were published after the review was conducted. In this case, we can safely assume that this threat is not so evident in our study since the \ac{SLR} was performed 5 months before its submission. 
\textbf{Publication Quality:} This occurs when studies of poor quality are included in the review. To mitigate this aspect, we defined inclusion criteria on the quality of the venue of the publication, leaving to a manual evaluation of the authors of this \ac{SLR} the inclusion of publications that appeared in venues of lower quality. 
\textbf{Grey Literature was not Considered:} This includes preprints and tools, such as commercial tools and platforms, that were not published in academic venues, despite their popularity or relevance to our RQs.

\revi{The organization and navigation criteria adopted in this \ac{SLR} should not be interpreted as exhaustive or definitive since they come from the authors' design choices. Alternative classification schemes or additional dimensions may be equally valid and could offer complementary insights, particularly for domain-specific or practice-oriented use cases. Moreover, the effectiveness and appropriateness of these navigation criteria have not yet been empirically evaluated with end users.}
\revi{To mitigate this limitation and enable stakeholders to customize their navigation and obtain insights into our results, we provide access to \ac{HERMES}. Additionally, while we acknowledge the importance of user-based validation of \ac{HERMES}, we consider it as a supporting and exploratory artifact within the broader scope of this work---which synthesizes and structures existing knowledge on \acp{XUI}---rather than as a standalone, fully validated system. \ac{HERMES} was in fact designed to complement this contribution by offering an interactive and actionable way to navigate the results of the \ac{SLR}, facilitating exploration by researchers, designers, and practitioners. A comprehensive user-centered evaluation of \ac{HERMES} (in particular assessing usability, usefulness, and impact on research and design practices) is therefore left as an important direction for future work.}
\revi{Finally, findings were reported by clusters to maintain contextual specificity and avoid overgeneralization across multiple domain contexts. This approach indeed limits the interpretational boundaries of the results since cross-domain transferability would require dedicated user studies, thus strengthening the methodological rigor of the \ac{SLR} and preventing misleading conclusions.}


\section{Conclusions}\label{sec:conclusions}
This study explores the design and evaluation of \ac{XUI}. While \acp{XUI} are widely recognised as a fundamental component of the explanation process, their implementation still presents significant research challenges. We approached this problem from a multidimensional perspective. On one hand, we examined \ac{HCI} requirements, analyzing user types and evaluation \revi{constructs} derived from user studies. On the other hand, we investigated algorithmic aspects, bridging these two dimensions through the interface layer. Our analysis led to the development of guidelines for designers, combining both top-down and bottom-up approaches. The top-down perspective involved reviewing studies that propose directives and frameworks, while the bottom-up approach focused on extracting design principles and requirements from case studies. These guidelines aim to close the gap between the usability of explainable interfaces and approaches that focus solely on algorithmic transparency. By providing a comprehensive view of the ecosystem in which \acp{XUI} are designed and implemented, this study helps identify key challenges in creating explainable processes that comply with emerging regulations. Ultimately, our findings contribute to making \ac{XAI} not only a research concept but a practical and deployable solution across real-world applications. In fact, in this work, we also introduced \ac{HERMES}, a platform for guiding practitioners in the design and evaluation of \acp{XUI}. Although user studies to evaluate and validate the platform are needed, with this platform, we attempt to bridge the gap that exists between results coming from the academic literature and their actual application by designers and developers.

%
\credit{EC}{Conceptualization, Data curation, Formal analysis, Investigation, Methodology, Validation, Visualization, Writing -- original draft, Writing -- review \& editing, Software}
\credit{AE}   {Data curation, Formal analysis, Investigation, Methodology, Validation, Visualization, Writing -- original draft, Writing -- review \& editing, Software}
\credit{FG}  {Data curation, Formal analysis, Investigation, Methodology, Validation, Visualization, Writing -- original draft, Writing -- review \& editing, Software}
\credit{GD} {Project administration, Funding acquisition, Supervision, Writing -- review \& editing}
\credit{RL} {Supervision, Funding acquisition, Project administration, Writing -- review \& editing}
\credit{SR}  {Project administration, Funding acquisition, Supervision, Writing -- review \& editing}

\paragraph{Authors Contribution Statement}
\insertcreditsstatement.

\bibliographystyle{ACM-Reference-Format}
\bibliography{sample-base,esposito-bibliography}
\end{document}